\documentclass[fdp,a4paper,fleqn]{w-art}
\usepackage{times,cite,w-thm}
\theoremstyle{plain}

\theoremstyle{definition}

\usepackage{graphicx}
\usepackage{amsmath}
\usepackage{amssymb}
\usepackage{mathrsfs}
\usepackage{dsfont}
\usepackage{epsfig}
\usepackage{graphicx}
\usepackage{subfigure}
\newcommand{\al}{\alpha}
\newcommand{\be}{\beta}

\newcommand{\mrm}[1]{\mathrm{#1}}

\newcommand{\LL}{\mathscr{L}}

\def\cO{{\cal O}}
\def\cK{{\cal K}}
\def\cW{{\cal W}}
\def\cM{{\cal M}}

\newcommand{\beq}{\begin{equation}}
\newcommand{\eeq}{\end{equation}}
\newcommand{\bac}{\beq\begin{array}}
\newcommand{\eac}{\end{array}\eeq}
\newcommand{\ba}{\begin{array}}
\newcommand{\ea}{\end{array}}
\newcommand{\bea}{\begin{eqnarray}}
\newcommand{\eea}{\end{eqnarray}}

\newcommand{\hc}{\text{h.c.}}
\newcommand{\unity}{\mathds{1}}
\newcommand\dd{\displaystyle}
\newcommand{\mean}[1]{\langle#1\rangle}

\newcommand{\ov}[1]{\overline{#1}}

\begin{document}
\DOIsuffix{theDOIsuffix}
\Volume{55}
\Month{01}
\Year{2007}
\pagespan{1}{}
\Receiveddate{XXXX}
\Reviseddate{XXXX}
\Accepteddate{XXXX}
\Dateposted{XXXX}

\title[$S_4$ Neutrino Models]{\boldmath Neutrino Mixings and the $S_4$ Discrete Flavour Symmetry}
\author[F. Bazzocchi]{Federica Bazzocchi\inst{1}}
\address[\inst{1}]{INFN -- Sezione di Trieste and
SISSA, via Bonomea 265, 34136 Trieste, Italy,}
\author[L. Merlo]{Luca Merlo\inst{2,3,}%
  \footnote{Corresponding author\quad E-mail:~\textsf{merlo@tum.de}}}
\address[\inst{2}]{Physik-Department, Technische Universit\"at M\"unchen, 
James-Franck-Strasse, D-85748 Garching, Germany -- preprint: TUM-HEP-834/12}
\address[\inst{3}]{TUM Institute for Advanced Study, Technische Universit\"at M\"unchen,
Lichtenbergstrasse 2a, D-85748 Garching, Germany}
\begin{abstract}
Discrete non-Abelian Symmetries have been extensively used to reproduce the lepton mixings. In particular, the $S_4$ group turned out to be suitable to describe predictive mixing patterns, such as the well-known Tri-Bimaximal and the Bimaximal schemes, which all represent possible first approximations of the experimental lepton mixing matrix. We review the main application of the $S_4$ discrete group as a flavour symmetry, first dealing with the formalism and later with the phenomenological implications. In particular, we summarize the main features of flavour models based on $S_4$, commenting on their ability in reproducing a reactor angle in agreement with the recent data and on their predictions for lepton flavour violating transitions.
\end{abstract}
\maketitle


\section{Introduction}
The presence of additional symmetries beyond the SM gauge group acting on the three fermion generations can produce realistic mass hierarchies and mixing textures. In such models, the Lagrangian is invariant under the gauge group of the SM and under the additional flavour symmetry at an energy scale $\Lambda_f$ equal or higher than the electroweak one. Fermion masses and mixings then arise once these symmetries are broken, spontaneously or explicitly. These flavour models differ from each other depending on the nature of the symmetries, either global or local, either Abelian or non-Abelian, either continuous or discrete, and on their symmetry breaking mechanism. Many examples have been produced in the last thirty years: at first only the quark sector has been investigated and by the implementation of the simple $U(1)$ \cite{Froggatt:1978nt} and $U(2)$ \cite{Pomarol:1995xc,Barbieri:1995uv,Barbieri:1996ww} symmetries; more recently, also the lepton sector received lot of attention, due to the more precise data on the neutrino oscillations \cite{Fogli:2012ua,Tortola:2012te}. In particular, discrete non-Abelian symmetries \cite{Altarelli:2010gt,Ishimori:2010au,Ludl:2010bj,Grimus:2010ak,Parattu:2010cy,Grimus:2011fk} have shown their ability in describing particular mixing patterns, that well describe the experimental data: these textures are the so-called Tri-Bimaximal \cite{Harrison:2002er,Harrison:2002kp,Xing:2002sw,Harrison:2002et,Harrison:2003aw} (TB), Bimaximal \cite{Vissani:1997pa,Barger:1998ta,Nomura:1998gm,Altarelli:1998sr} (BM), Golden Ratio \cite{Kajiyama:2007gx,Rodejohann:2008ir,Adulpravitchai:2009bg,Everett:2008et,Feruglio:2011qq,Ding:2011cm} (GR) and Trimaxinal \cite{King:2011zj}(TM) schemes.

Apart from the symmetries, flavour models differ from each other depending on the choice of the flavour symmetry breaking scale, that could either correspond to the electroweak scale or be larger, up to the grand unification scale. In the first case, replicants of the SM Higgs, with non-trivially transformation properties under the flavour symmetry, usually enrich the spectrum.  These models could suffer of strong constraints from Flavour Changing Neutral Current (FCNC) processes, that receive contributions from Higgs exchanges \cite{Toorop:2010ex,Toorop:2010kt,Machado:2010uc,Cao:2011df,Branco:2011iw,deAdelhartToorop:2011ad}. In the second case, new scalar fields, invariant under the SM gauge group but transforming under the flavour symmetry, enrich the spectrum of the models. Their masses and vacuum expectation values (VEVs) are typically very large, close to the grand unification energy scale, and as a result they have negligible direct contributions to FCNC processes. Indirect contributions, however, are possible in beyond SM contexts with the presence of new physics (NP) \cite{Calibbi:2012at}, such as supersymmetry \cite{Hamaguchi:2002vi,Mondragon:2007af,Kifune:2007fj,Ishimori:2008ns,Feruglio:2008ht,Ishimori:2008au,Feruglio:2009iu,Feruglio:2009hu,Hagedorn:2009df,Merlo:2011hw,Chakrabortty:2012vp} or extra dimensions (ED) \cite{Perez:2008ee,Csaki:2008qq}.

In order to avoid large contributions to FCNC processes, it could be useful to follow a bottom-up approach: it consists first in identifying low-energy effective schemes in which the contributions to FCNC observables are under control and subsequently in constructing high-energy models which could origin such effective descriptions. The so-called Minimal Flavour Violation scheme (MFV) \cite{Chivukula:1987py,Hall:1990ac} follows this approach. Since no evident (larger than $3\sigma$'s) deviations from the SM predictions have been found in all flavour processes observed in the hadronic sector, from rare decays in the kaon and pion sectors to $B$ decays at superB--factories \cite{Isidori:2010kg}, it seems reasonable that any physics beyond the SM does not introduce new sources of flavour and CP violation with respect to the SM. In Refs.~\cite{D'Ambrosio:2002ex,Cirigliano:2005ck,Davidson:2006bd,Grinstein:2006cg,Alonso:2011jd} this criterion has been rigorously defined in terms of continuous flavour symmetries, considering an effective operator description in the SM context. Even if the MFV ansatz is successful to prevent large flavour violation also in models beyond SM \cite{Grinstein:2010ve,Feldmann:2010yp,Guadagnoli:2011id,Buras:2011zb,Arcadi:2011ug,Buras:2011wi,Alonso:2012jc}, it does not provide any explanation for the origin of fermion masses and mixings. From this the idea that the MFV is a too restrictive context to provide any clue on the origin of masses. Some extensions of the MFV where smaller continuous symmetries have been adopted can be found in Refs.~\cite{Barbieri:2011ci,Crivellin:2011sj,Barbieri:2011fc}, but also in these analyses a complete explanation of fermion masses is missing.

In this review we deal with discrete non-Abelian symmetries that lead to a more complete explanation of the fermion masses. Furthermore, we restrict our analysis only on models where the flavour symmetry is broken at a high scale, that is a more promising framework to prevent large flavour violation. More in details, we concentrate on the $S_4$ discrete group that has been adopted to describe the TB, BM and TM neutrino patterns. In Sect.~\ref{sec:NuPatterns}, we motivate the choice of discrete symmetries to handle the neutrino sector and describe the aforementioned mixing schemes, commenting on their agreement with the data. In Sect.~\ref{sec:group}, we describe the $S_4$ group and three relevant bases, that have been used in the literature. In Sect.~\ref{sec:gen}, we summarize the main features of flavour models dealing with the $S_4$ group, including the extension to the quark sector. In Sect.~\ref{sec:Pheno}, we study some phenomenological implications of the $S_4$ models, such as the presence of neutrino mass sum-rules, the prediction for the neutrino-less-double-beta ($0\nu2\beta$) decay effective mass and the constraints coming from lepton flavour violating decays. Finally, in Sect.~\ref{sec:Concl} we conclude.

\section{Lepton Mixing Patterns}
\label{sec:NuPatterns}

The \emph{precision data} era for lepton mixing started only ten years ago but since then our knowledge of lepton mixing angles as well as that of neutrino mass splitting have grown exponentially.
Nowadays, the most recent results on the oscillation data analysis\footnote{Notice that the best fit values for the reactor angle differ for a factor of 2 in the two fits, due to the exclusion (Fogli {\it et al.}) or the inclusion (Schwetz {\it et al.}) of the data from SBL neutrino experiments with a baseline $<100$ m.} can be summarized in Tab.~\ref{tab:data}.

\begin{vchtable}[h]
\vchcaption{Recent fits to neutrino oscillation data from \cite{Fogli:2012ua,Tortola:2012te}. In the brackets the IH case. $)^\star$ In this case the full $(0,\,2\pi)$ is allowed.}
\label{tab:data}
\begin{tabular}{@{}ccc@{}}
  \hline
  &&\\[-3mm]
  Quantity & Fogli {\it et al.} \cite{Fogli:2012ua} & Schwetz {\it et al.} \cite{Tortola:2012te} \\[1mm]
  \hline
  &&\\[-3mm]
  $\Delta m^2_{sun}~(10^{-5}~{\rm eV}^2)$ 			&$7.54^{+0.26}_{-0.22}$ 																																								&$7.62\pm0.19$  \\[1mm]
  $\Delta m^2_{atm}~(10^{-3}~{\rm eV}^2)$ 			&$2.43^{+0.07}_{-0.09}$ ($2.42^{+0.07}_{-0.1}$) 																															&$2.53^{+0.08}_{-0.10}$ ($2.40^{+0.10}_{-0.07}$)  \\[1mm]
  $\sin^2\theta_{12}$ 											&$0.307^{+0.018}_{-0.016}$ 																																							&$0.320^{+0.015}_{-0.017}$ \\[1mm]
  $\sin^2\theta_{23}$ 											&$0.398^{+0.03}_{-0.026}$ ($0408^{+0.035}_{-0.03}$) 																													&$0.49^{+0.08}_{-0.05}$ ($0.53^{+0.05}_{-0.07}$) \\[1mm]
  $\sin^2\theta_{13}$ 											&$0.0245^{+0.0034}_{-0.0031}$ ($0.0246^{+0.0034}_{-0.0031}$) 																									&$0.026^{+0.003}_{-0.004}$ ($0.027^{+0.003}_{-0.004}$)  \\[1mm]
  $\delta_{CP}/\pi$ 												&$0.89^{+0.29}_{-0.44}$ ($0.90^{+0.32}_{-0.43}$)																															&$0.83^{+0.54}_{-0.64}$ ($0.07^\star$)\\
  \hline
\end{tabular}
\end{vchtable}

The past and present years were important for neutrino physics because in the summer 2011 the last T2K data \cite{Abe:2011sj} showed evidences for a non-vanishing reactor angle at the $3\sigma$ level. Subsequently also MINOS \cite{Adamson:2011qu} and Double Chooz \cite{Abe:2011fz} presented their results, in agreement with T2K one. More recently the Daya Bay \cite{An:2012eh} and RENO \cite{Ahn:2012nd} experiments have released their results on the observation of electron anti-neutrino disappearance, providing at more than $5\sigma$ and $6\sigma$, respectively, the evidence for a non-vanishing reactor angle:
\beq
\sin^2\theta_{13}=0.024 \pm 0.005\qquad \text{[Daya Bay]}\,,\qquad\qquad
\sin^2\theta_{13}=0.029 \pm 0.006\qquad \text{[RENO]}\,.
\eeq
The average of the three most precise results for the reactor angle from DOUBLE CHOOZ\cite{Abe:2011fz}, Daya Bay \cite{An:2012eh} and RENO \cite{Ahn:2012nd} experiments is given by
\beq
\sin^2\theta_{13}=0.0253\pm0.0035\,,
\label{OurReactor}
\eeq
for both the mass orderings.

These results indicate the double nature of \emph{neutrino puzzle}: indeed, the neutrino mass scale and mass hierarchies are smaller of many orders of magnitude with respect to those of the other fermions and the lepton mixing is completely different from quark mixing, where the largest mixing angle is the Cabibbo angle $\lambda=\sin\theta_C\simeq0.22$. For these reasons neutrino physics has raised so much interest in the last years. In this review we will focus on the origin  of  lepton mixing and will not address the other interesting question related to the origin of their masses. In all the models we will present neutrino masses are obtained by means of See-Saw mechanisms or more simply by means of the Weinberg operator.

Without any kind of prejudice we may state that there are two motivations in support of the idea that lepton mixing is ascribed to the breaking of non-Abelian discrete flavour symmetry. Firstly, the presence of two large angles have suggested the idea that the lepton mixing matrix cannot be well described by mass ratios, as for the CKM matrix. In order to encode this ansatz, in Ref.~\cite{Low:2003dz} the concept of Form Diagonalizable (FD) matrices has been introduced: that is mass matrices that are diagonalized by unitary matrices whose entries are independent of the mass eigenvalues. In the following, with a small abuse of notation, we will refer to these unitary matrices as FD mixing matrices even if the FD acronym refers only to the mass matrices. Secondly, under the assumption that neutrinos are majorana particles, if there is any residual symmetry behind neutrino mass matrix this is at most a $Z_2 \times Z_2$  flavour symmetry \cite{Toorop:2011jn,deAdelhartToorop:2011re}.  The technical implication of this feature will be seen in next sections. In the rest of the section, we concentrate on the predictive FD mixing patterns.\\ 

First of all we remind the standard parametrization adopted for lepton mixing. This is  written in terms of 3 mixing angles, 1 Dirac phases and 2  Majorana phases. The three mixing angles, $\theta_{12}$, $\theta_{23}$, $\theta_{13}$ are also denoted as solar, atmospheric and reactor angles in order to remember the kind of neutrino source. The standard form is given by
\beq
\label{parmix}
\begin{array}{ll}
U_{lep}&= R_{23} (\theta_{23})\cdot R_{13} (\theta_{13},\delta)\cdot R_{12} (\theta_{12}) \cdot P_\phi\\
&\\
&= \left( \begin{array}{ccc} 1 &0&0\\ 0& c_{23}&s_{23}\\ 0& -s_{23}& c_{23} \end{array} \right)\cdot  \left( \begin{array}{ccc} c_{13} &0&s_{13} e^{i \delta} \\ 0& 1&0\\ -s_{13} e^{-i \delta}&0& c_{13} \end{array} \right) \cdot  \left( \begin{array}{ccc} c_{12} &s_{12} &0 \\ -s_{12}& c_{12}&0\\ 0&0&1 \end{array} \right)\cdot P_\phi
\end{array}
\eeq
with $R_{ij} (\theta_{ij})$ a rotation in the $i-j$ plane of an angle $\theta_{ij}$, $P_\phi=\mbox{diag}(1, e^{i \alpha_{21}},e^{i \alpha_{31}})$, where $\alpha_k$ are the Majorana phases, and $c_{ij},s_{ij}= \cos{\theta_{ij}},\sin{\theta_{ij}}$. According to Ref.~\cite{Low:2003dz}, a FD mixing pattern is characterized by having all the $c_{ij},s_{ij}$ as pure numbers, in particular independent from the mass eigenstates. For this reason a specific FD mixing pattern is not correlated neither to the neutrino spectrum nor to the Majorana phases. In concrete models the constraints on the neutrino mass splittings typically provide specific ranges for the Majorana phases. More these ranges are small more predictive, and thus interesting, are the models. 

The first FD mixing pattern that we consider was suggested by Harrison Perkins Scott in 2002 \cite{Harrison:2002er,Harrison:2002kp,Xing:2002sw,Harrison:2002et,Harrison:2003aw} and is now referred as the Tri-Bimaximal (TB) scheme:
\bea
U_{TB}&=&\left (\displaystyle{ \begin{array}{ccc} \frac{2}{\sqrt{6}} & \frac{1}{\sqrt{3}} & 0\\
-\frac{1}{\sqrt{6}} & \frac{1}{\sqrt{3}} & -\frac{1}{\sqrt{2}}\\
-\frac{1}{\sqrt{6}} & \frac{1}{\sqrt{3}} & \frac{1}{\sqrt{2}}
\end{array}}\right) \,,
\eea
that corresponds to the following mixing angles
\beq
\sin\theta_{23}^2 =\frac{1}{2}\,,\quad \sin\theta_{12}^2 =\frac{1}{3}\,,\quad \sin\theta_{13}^2 =0\,.
\eeq
The most general mass matrix diagonalized by the TB mixing is given by
\begin{equation}
m_\nu^{TB}=\left(\begin{array}{ccc}
x&y&y\\
y&x+z&y-z\\
y&y-z&x+z
\end{array}\right)\,,
\label{MMTB}
\end{equation}
with complex parameters $x$, $y$ and $z$. This matrix is invariant under the $\mu-\tau$ symmetry, i.e. the simultaneous exchange of the second and third rows and columns, and under the so-called magic symmetry, for which the $22$ and $23$ entries are written in these particular combinations of $x$, $y$ and $z$.

The second FD mixing pattern that we present is the Bimaximal (BM) scheme \cite{Vissani:1997pa,Barger:1998ta,Nomura:1998gm,Altarelli:1998sr}. Historically the BM scenario for lepton mixing was proposed before the TB one, but not in the context of discrete non-Abelian flavour symmetry and for this reason in our review comes as second. It is defined by
\bea
\label{genBM}
U_{BM}&=&\left (\displaystyle{ \begin{array}{ccc} \frac{1}{\sqrt{2}} & -\frac{1}{\sqrt{2}} & 0\\
\frac{1}{2} & \frac{1}{2} & -\frac{1}{\sqrt{2}}\\
\frac{1}{2} & \frac{1}{2} & \frac{1}{\sqrt{2}}
\end{array}}\right) \,,
\eea
corresponding to the following mixing angles
\beq
 \sin\theta_{23}^2 =\frac{1}{2}\,,\quad \sin\theta_{12}^2 =\frac{1}{2}\,,\quad \sin\theta_{13}^2 =0\,.
\eeq
In this case, the most general mass matrix of the BM type is given by
\begin{equation}
m_\nu^{BM}=\left(\begin{array}{ccc}
x&y&y\\
y&x+z&-z\\
y&-z&x+z
\end{array}\right)\,,
\label{MMBM}
\end{equation}
with complex parameters $x$, $y$ and $z$. This mass matrix is still invariant under the $\mu-\tau$ symmetry and under an equivalent version of the magic symmetry, that defines the $22$ and $23$ entries.

Both TB and BM mixing predicts at leading order (LO) a vanishing reactor angle. However, the new data on the reactor angle have motivated the analytical research of new mixing pattern that could describe, at LO, lepton mixings with a non-vanishing reactor angle. The most complete analysis of this kind have been performed in \cite{Toorop:2011jn,deAdelhartToorop:2011re}. However $S_4$ may not been used to predict such patterns\cite{deAdelhartToorop:2011re} and in the following we will only discuss suitable modifications of the TB and BM patterns.

There is a substantial historically difference between the TB and the BM mixing patterns: when TB scheme was introduced it was in excellent agreement with experimental data. As a result, a model that predicts the exact TB patter as the lepton mixing did not need any corrections, from the phenomenological point of view. However in any model so far studied, even in those ultraviolate (UV) completed \cite{Varzielas:2010mp}, sub-leading corrections to lepton mixing are always present. On the contrary, the solar angle of the BM pattern has never been in agreement with the experimental data fit: the mixing angles are fitted thanks to a relative large correction to the solar angle, of order of the Cabibbo angle $\lambda$, while only minor corrections affect the other two angles. This makes the BM pattern highly interesting to revisit the idea of quark-lepton complementarity \cite{Altarelli:2004jb,Raidal:2004iw,Minakata:2004xt,Frampton:2004vw,Ferrandis:2004vp}.

With the new results on the reactor angle, both TB and BM mixing scheme need corrections to fit the data. The origin of such corrections is model dependent  but we may catalogue them into two big  groups once we go in the basis in which all the FD structure arise from the neutrinos: they may arise from the diagonalization of either the charged lepton mass matrix or the neutrino mass matrix, once the sub-leading corrections are taken into account. Clearly in the most generic case the corrections have both origins.

Suppose that at LO in the basis in which the charged lepton mass matrix $M^{(0)}_e$ is diagonal the neutrino mass matrix $M^{(0)}_\nu$ is diagonalized by the FD mixing matrix $U_{FD}$. Thus at NLO we said that we may have the following scenarios:
\begin{description}
\item[1) Corrections only from the charged lepton sector] 
\beq
M_e =M_e^{(0)}+\delta M_e\,,\qquad\qquad
M_\nu =M_\nu^{(0)}\,.
\eeq
In this way $M_e M_e^\dag$ is diagonalized by a unitary matrix $V_e$ that may be written as
\beq
\label{vl}
V_e=\left (\displaystyle{ 
\begin{array}{ccc}
1& c^e_{12}\,\theta & c^e_{13}\, \theta \\
-c^{e*}_{12}\, \theta & 1 & c^e_{23}\,\theta\\
-c^{e*}_{13}\, \theta & -c^{e*}_{23}\, \theta &1
\end{array}}\right) \,,
\eeq
where  $c^e_{12}$, $c^e_{13}$, $c^e_{23}$ are in the most general case complex parameters with absolute value of order 1.
According to the TB and BM schemes the final lepton mixing matrix is given by
\beq
U_{lep}=V_e^\dag\, U_{FD},
\eeq
where FD stays for TB or BM schemes. The corresponding lepton mixing angles for the TB pattern are given by 
\beq
\label{newangTB}
\begin{aligned}
(\sin\theta^{2}_{23})_{TB}=&\frac{1}{2}+{\cal R}e(c^e_{23})\,\theta\,,\\
( \sin\theta^{2}_{12})_{TB} =&\frac{1}{3}-\frac{2}{3}{\cal R}e\left(c^e_{12}+c^e_{13}\right)\,\theta\,,\\
( \sin\theta_{13})_{TB} =&\frac{1}{\sqrt2}\left|c^e_{12}-c^e_{13}\right|\,\theta\,,
 \end{aligned}
\eeq
while for the BM case by
\beq
\label{newangBM}
\begin{aligned}
(\sin\theta^2_{23})_{BM} =&\frac{1}{2}+{\cal R}e(c^e_{23})\,\theta\,,\\
(\sin\theta^2_{12})_{BM} =&\frac{1}{2}+\frac{1}{\sqrt{2}}{\cal R}e(c^e_{12}+c^e_{13})\,\theta\,,\\
(\sin\theta_{13})_{BM} =&\frac{1}{\sqrt2}\left|c^e_{13}-c^e_{12}\right|\theta\,.
\end{aligned}
\eeq

By looking at eqs.~(\ref{newangTB}-\ref{newangBM})  for any  $\theta$ we may wonder  which is the probability to have all the three angles in the $3\sigma$ ranges allowed by the data assuming the absolute values of the $c^e_{12}$, $c^e_{23}$ and $c^e_{13}$ are distributed as  gaussian centered in one --to be consistent with the assumption they are of order one. By plotting such probability as function of $\theta$ we get the Success Rate (SR) of the scheme as shown in Fig.\ref{Succ-ch}. The $\theta$ value that maximizes the success rate, $\theta_{SR}$, may be seen has the most natural value for $\theta$ to fit the data. By fixing  $\theta=\theta_{SR}$ we may visualize the possible correlations among the lepton mixing  angles by means of scatterplots: for example by looking at  eqs.~(\ref{newangTB})-(\ref{newangBM})  we see that $\theta_{23}$ is expected to be only mildly correlated with the other two angles while there could be a  correlation between the solar and the reactor angles. However if we take the absolute value of the parameters $c^e_{ij}$ according to a normal distribution centered in $\pm 1$ the correlation almost disappears as shown in Fig.~\ref{Scatch}. 

\begin{figure}[h!]
  \centering
  \includegraphics[width=7.4cm]{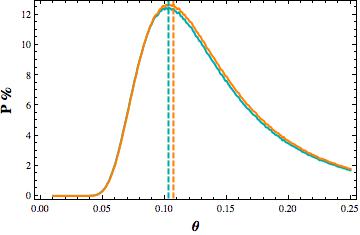}
  \includegraphics[width=7.4cm]{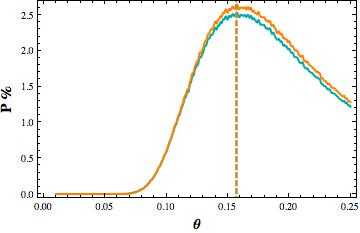}
\caption{SRs for the TB (left panel) and BM  (right panel) schemes  at NLO for the NH (cyan) and IH (orange) cases according to the parametrization done in  eqs.\ref{newangTB}-\ref{newangBM}  assuming that the corrections arise only by the charged leptons. The SRs have been obtained by discretizing the interval $(0.01,0.25)$ for $\theta$ and by taking randomly $10^6$ sets $(c^e_{12},c^e_{13},c^e_{23})$ for each bin. The $c^e_{ij}$ have been generated assuming a random phase and an absolute value normal distributed around 1 with $\sigma=0.5$. The $\theta$ values that maximizes the SRs are for both the hierarchies $0.103$  and $0.107$ for the TB and BM schemes, respectively.}
\label{Succ-ch} 
\end{figure}

\begin{figure}[h!]
  \centering
  \includegraphics[width=7.4cm]{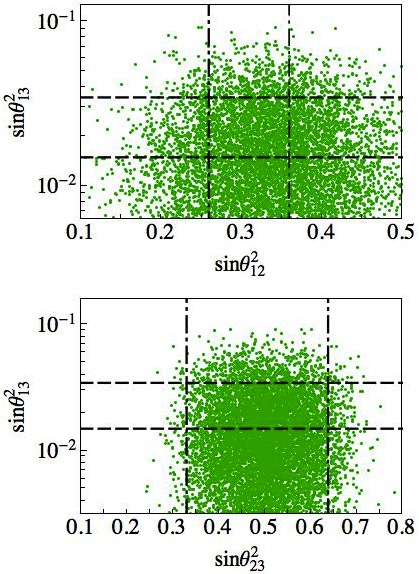}
   \includegraphics[width=7.4cm]{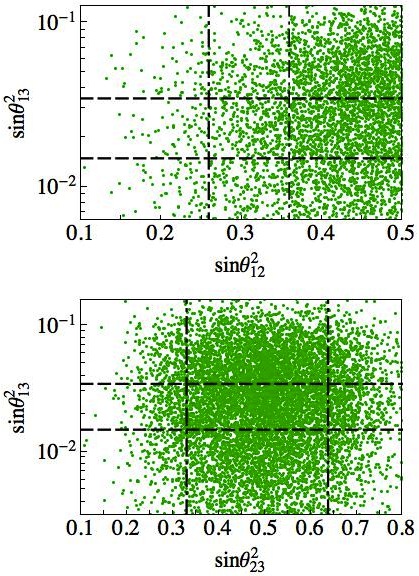}
\caption{On the upper (lower) line  the reactor angle versus the solar (atmospheric) angle for the TB, on the left, and the BM, on the right, schemes at NLO according to the parametrization done in  eqs.\ref{newangTB}-\ref{newangBM}, assuming that the corrections arise by the charged lepton sector. The scatter plots have been obtained by  fixing $\theta=\theta_{SR}$ being $\theta_{SR}$ the value that maximizes the SR according to Fig.\ref{Succ-ch} and by taking the $|c^e_{ij}| $ centered in 1 with normal distribution. Horizontal and vertical lines bound the updated $3\sigma$ ranges for the mixing angles, following the most recent global fit \cite{Fogli:2012ua}. }
\label{Scatch}
\end{figure}

\item[2) Corrections only from the neutrino sector] 
\beq
M_e =M_e^{(0)}\,,\qquad\qquad
M_\nu =M_\nu^{(0)}+\delta M_\nu\,.
\eeq
In this case, $M_{\nu}^{(0)}$ is diagonalized by $U_{FD}$, according to $\hat{M}_\nu^{(0)}= U_{FD}^T\cdot M_\nu^{(0)}\cdot U_{FD}$, where $\hat{M}_\nu^{(0)}$ is a diagonal matrix. In this basis, $\delta\hat{M}_{\nu}$ can be parametrized by
\beq
\delta\hat{M}_{\nu}=\left (\begin{array}{ccc} 0 & \epsilon_1 & \epsilon_2\\
\epsilon_1 & 0 & \epsilon_3\\
\epsilon_2& \epsilon_3 &0
\end{array}\right) \,,
\eeq
where the diagonal corrections have been absorbed redefining  the diagonal entries of \mbox{$\hat{M}_{\nu}^{(0)}= {\rm diag} (m_1,m_2,m_3)$}. In this basis the full neutrino mass matrix 
\beq
\hat{M}_\nu= \hat{M}_{\nu}^{(0)}+ \delta\hat{M}_{\nu}\,,
\eeq
is diagonalized by a unitary matrix $V_\nu$, 
\beq
\label{vnu}
V_\nu=\left (\displaystyle{ 
\begin{array}{ccc}
1& c^\nu_{12}\,\theta & c^\nu_{13}\, \theta \\
-c^{\nu*}_{12}\, \theta & 1 & c^\nu_{23}\,\theta\\
-c^{\nu*}_{13}\, \theta & -c^{\nu*}_{23}\, \theta &1
\end{array}}\right) \,,
\eeq
with the $|c^\nu_{ij}|\sim \mathcal{O}(1)$ and $|\theta|<1$.  By writing the $\epsilon_i= \hat{\epsilon}_i\, \theta$ we easily get
\beq
|c^\nu_{12}|\simeq \frac{|\hat{\epsilon}_1|}{m_2-m_1}\,,\quad 
|c^\nu_{13}|\simeq \frac{|\hat{\epsilon}_2|}{m_3-m_1}\,,\quad 
|c^\nu_{23}|\simeq \frac{|\hat{\epsilon}_3|}{m_3-m_2}\,.
\eeq
The expression for the mixing angles in the TB case are given by
\beq
\label{newangTBb}
\begin{aligned}
(\sin\theta^{2}_{23})_{TB}=&\frac{1}{2}+\dfrac{1}{\sqrt{3}}{\cal R}e\left(c^\nu_{13}-\sqrt2\,c^\nu_{23}\right)\,\theta\,,\\
( \sin\theta^{2}_{12})_{TB} =&\frac{1}{3}+\dfrac{2\sqrt2}{3}\,{\cal R}e(c^\nu_{12})\,\theta\,,\\
( \sin\theta_{13})_{TB} =&\frac{1}{\sqrt3}\left|\sqrt2\,c^\nu_{13}+c^\nu_{23}\right|\,\theta\,,
 \end{aligned}
\eeq
while for the BM case by
\beq
\label{newangBMb}
\begin{aligned}
(\sin\theta^2_{23})_{BM} =&\frac{1}{2}-\dfrac{1}{\sqrt{2}}{\cal R}e\left(c^\nu_{13}+c^\nu_{23}\right)\,\theta\,,\\
( \sin\theta^2_{12})_{BM} =&\frac{1}{2}-{\cal R}e(c^\nu_{12})\,\theta\\
( \sin\theta_{13})_{BM} =&\frac{1}{\sqrt2}\left|c^\nu_{13}-c^\nu_{23}\right|\theta\,,
 \end{aligned}
\eeq
The corresponding SRs are shown in Fig. \ref{Succ-co}.

\begin{figure}
  \centering
  \includegraphics[width=7.4cm]{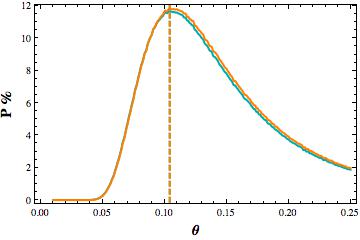}
   \includegraphics[width=7.4cm]{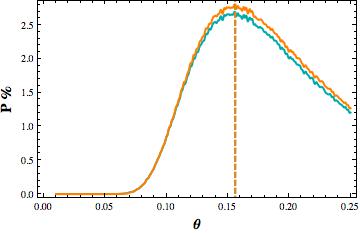}
  \caption{SRs for the TB (left panel) and BM  (right panel) schemes  at NLO  for the NH (cyan) and IH (orange) cases according to the parametrization done in  eqs.\ref{newangTBb}-\ref{newangBMb},  assuming that the corrections arise  only by the neutrino mass matrix. The $\theta$ values that maximize the SRs are for both the hierarchies $0.1$ and $0.15$ for the TB and BM case, respectively. The SRs have been obtained as in Fig.\ref{Succ-ch}.}
\label{Succ-co}
\end{figure}

\begin{figure}
  \centering
  \includegraphics[width=7.4cm]{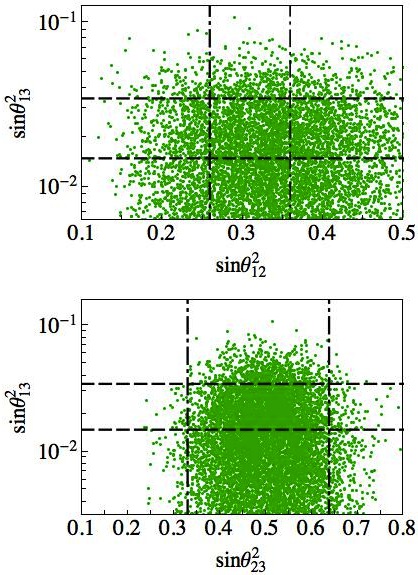}
   \includegraphics[width=7.4cm]{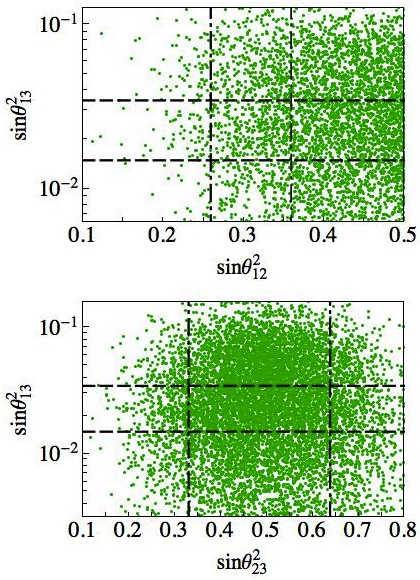}
  \caption{ On the upper (lower) line the reactor angle  versus the solar(atmospheric) angle for the TB, on the left, and BM, on the right, schemes at NLO according to the parametrization done in  eqs.~(\ref{newangTBb})-(\ref{newangBMb}),  assuming that the corrections to the mixing  angles arise by the neutrino mass matrix.  The scatter  plots have been obtained by  fixing $\theta=\theta_{SR}$ being $\theta_{SR}$ the value that maximizes the corresponding SR according to Fig.~\ref{Succ-co} and by taking the $|c^\nu_{ij}| $ centered in 1 with normal distribution. Horizontal and vertical lines bound the updated $3\sigma$ ranges for the mixing angles, following the most recent global fit \cite{Fogli:2012ua}. }
\label{Scatco}
\end{figure}
\end{description}

As previously noticed, we remind that in general both kind of corrections are present and thus some correlations are weakened. However in realistic models more constraints are imposed, as for example those relative to  neutrino mass splittings and charged lepton masses. In this way  the models may maintain a good level of predictivity.
 
As a conclusion, according to our parametrization and the results reported in Figs.~\ref{Succ-ch} and \ref{Succ-co}, the BM scheme with corrections arising from either the charged lepton sector or the neutrino sector is strongly disfavoured, while, on the other side, the TB scheme can be characterized by a success rate up to the $13\%$, with a slightly best result when the corrections arise from the charged lepton sector. 
These results are in agreement with the analysis presented in Ref.~\cite{Altarelli:2012bn}. However, in concrete model realizations it is possible that the corrections affecting the lepton mixing matrix are not generic and as a result the success rate could reach much higher values: an example we could refer to the model in Ref.~\cite{Lin:2009bw}, that has been deeply analyzed in Ref.~\cite{Altarelli:2012bn}. 

\boldmath
\section{The Group $S_4$}
\label{sec:group}
\unboldmath

The group $S_4$ is the permutation group of four distinct objects, isomorphic to the group $O$ which is the symmetry group of a regular octahedron.  It may be thought as generated by two generators $S,T$, satisfying 
\beq
S^4=T^3= E\,, \qquad\qquad S T^2 S= T\,,
\eeq
being $E$ the neutral element of the group. 

It has $4!= 24$ distinct elements filled in five conjugate classes according to
\beq
\begin{array}{lll}
\mathcal{C}_1&:& E\,;\\
\mathcal{C}_2&:& S^2, TS^2 T^2,S^2 TS^2 T^2\,;\\
\mathcal{C}_3&:& T,T^2,S^2 T, S^2 T^2, S T ST^2,STS,S^2 TS^2 S^3 T S\,;\\
\mathcal{C}_4&:& S T^2,T^2 S,TST,TSTS^2,STS^2,S^2 TS\,;\\
\mathcal{C}_5&:& S,TST^2,ST,TS,S^3,S^3 T^2\,.\\
\end{array}
\eeq
As consequence there are five  irreducible representations that can be read from the character table reported in Tab.~\ref{tabcht}:  two   one-dimensional and   denoted as $1_1$ and $1_2$, one two-dimensional labelled as $2$ and two three-dimensional written as $3_1$ and $3_2$. 

\begin{vchtable}[h!]
\vchcaption{Character table for $S_4$. $\chi^r$ are the five representations and $\mathcal{C}_i$ the five conjugacy classes, $n$ the number of elements for each class and $h$ the smallest value for which $(\chi^{r})^h= 1$.}
\label{tabcht}
\begin{tabular}{@{}cccccccc@{}}
\hline
&&&&&&&\\[-3mm]
Class &$n$& $h$&$\chi^{1_1}$& $\chi^{1_2}$& $\chi^{2}$& $\chi^{3_1}$& $\chi^{3_2}$ \\[1mm]
\hline
&&&&&&&\\[-2mm]
$\mathcal{C}_1$ &1&1&1&1&2&3&3\\
$\mathcal{C}_2$ &3&2&1&1&2&-1&-1\\
$\mathcal{C}_3$ &8&3&1&1&-1&0&0\\
$\mathcal{C}_4$ &6&2&1&-1&0&1&-1\\
$\mathcal{C}_5$&6&4 &1&-1&0&-1&1\\[1mm]
\hline
\end{tabular}
\end{vchtable}

The multiplication rules are given by
\beq
\begin{array}{l}
1_1\times R=R\times1_1=R\qquad \text{where $R$ stands for any representation}\\
1_2\times1_2=1_1\\
1_2\times2=2\\
1_2\times3_1=3_2\\
1_2\times3_2=3_1\\
\\
2\times2=1_1+1_2+2\\
2\times3_1=3_1+3_2\\
2\times3_2=3_1+3_2\\
\\
3_1\times3_1=3_2\times3_2=1_1+2+3_1+3_2\\
3_1\times3_2=1_2+2+3_1+3_2\;.
\end{array}
\eeq

A possible choice for the generators $S$ and $T$ is given by
\beq
\begin{aligned}
&S_{1_1,1_2}=\pm1 \qquad 
&&T_1=1\\
&S_2=\left( \begin{array}{cc} -1&0\\0&1\end{array}\right)\qquad 
&&T_2=-\frac{1}{2}\left( \begin{array}{cc} 1&\sqrt{3}\\-\sqrt{3}&1\end{array}\right)\\
&S_{3_1,3_2}=\pm\left( \begin{array}{ccc} -1&0&0\\0&0&-1\\0&1&0\end{array}\right)\qquad 
&&T_3=\left( \begin{array}{ccc} 0&0&1\\1&0&0\\0&1&0\end{array}\right)\,,
\end{aligned}
\eeq
where  $X_r= r(X)$ being $r$ the representation of the generator $X$ and where we omitted the double index for $T$ since the representation is identical for both the singlets (triplets). 

With the previous choice of the generators is easy to extract the Clebsch-Gordan (CG) coefficients. In the following we use $\alpha_i$ to indicate the elements of the first representation of the product and $\beta_i$ to indicate those of the second one.
The multiplication rules  for the $1$-dimensional representations are given by:
\beq
\begin{aligned}
&1_1\otimes\eta&&=&&\eta\otimes1_1\,\,=\,\,\eta\qquad\text{with $\eta$ any representation}\\
&1_2\otimes1_2&&=&&1_1\sim\alpha\beta\\
&1_2\otimes2&&=&&2\sim\left(\begin{array}{c}
                    -\alpha\beta_2 \\
                    \alpha\beta_1 \\
            \end{array}\right)\\
&1_2\otimes3_1&&=&&3_2\sim\left(\begin{array}{c}
                    \alpha\beta_1 \\
                    \alpha\beta_2 \\
                    \alpha\beta_3 \\
                    \end{array}\right)\\
&1_2\otimes3_2&&=&&3_1\sim\left(\begin{array}{c}
                            \alpha\beta_1 \\
                            \alpha\beta_2 \\
                            \alpha\beta_3 \\
                    \end{array}\right)
\end{aligned}
\eeq
The multiplication rules with the 2-dimensional representation are the following:
\beq
\begin{aligned}
&2\otimes2=&&1_1\oplus1_2\oplus2\qquad&&
\text{with}\quad\left\{\begin{array}{l}
                    1_1\sim\alpha_1\beta_1+\alpha_2\beta_2\\[-10pt]
                    \\[8pt]
                    1_2\sim-\alpha_1\beta_2+\alpha_2\beta_1\\[-10pt]
                    \\[8pt]
                    2\sim\left(\begin{array}{c}
                        \alpha_1\beta_2+ \alpha_2\beta_1 \\
                        \alpha_1\beta_1-\alpha_2\beta_2 \\
                    \end{array}\right)
                    \end{array}
            \right.\\
&2\otimes3_1=&&3_1\oplus3_2\qquad&&
\text{with}\quad\left\{\begin{array}{l}
                    3_1\sim\left(\begin{array}{c}
                        \alpha_2\beta_1 \\
                        -\frac{1}{2}(\sqrt{3}\alpha_1\beta_2+\alpha_2\beta_2) \\
                       \frac{1}{2}(\sqrt{3}\alpha_1\beta_3-\alpha_2\beta_3) \\
                    \end{array}\right)\\[-10pt]
                    \\[8pt]
                    3_2\sim\left(\begin{array}{c}
                         \alpha_1\beta_1 \\
                        \frac{1}{2}(\sqrt{3}\alpha_2\beta_2-\alpha_1\beta_2) \\
                       -\frac{1}{2}(\sqrt{3}\alpha_2\beta_3+\alpha_1\beta_3) \\
                    \end{array}\right)\\
                    \end{array}
            \right.\\
&2\otimes3_2=&&3_1\oplus3_2\qquad&&
\text{with}\quad\left\{\begin{array}{l}
                    3_1\sim\left(\begin{array}{c}
                        \alpha_1\beta_1 \\
                        \frac{1}{2}(\sqrt{3}\alpha_2\beta_2-\alpha_1\beta_2) \\
                       -\frac{1}{2}(\sqrt{3}\alpha_2\beta_3+\alpha_1\beta_3) \\
                    \end{array}\right)\\[-10pt]
                    \\[8pt]
                    3_2\sim\left(\begin{array}{c}
                         \alpha_2\beta_1 \\
                        -\frac{1}{2}(\sqrt{3}\alpha_1\beta_2+\alpha_2\beta_2) \\
                       \frac{1}{2}(\sqrt{3}\alpha_1\beta_3-\alpha_2\beta_3) \\
                    \end{array}\right)\\
                    \end{array}
            \right.
\end{aligned}
\eeq
The multiplication rules with the 3-dimensional representations are the following:
\beq
\begin{aligned}
&3_1\otimes3_1=&&3_2\otimes3_2\,\,=\,\,1_1\oplus2\oplus3_1\oplus3_2&&\qquad
\text{with}\quad\left\{
\begin{array}{l}
1_1\sim\alpha_1\beta_1+\alpha_2\beta_2+\alpha_3\beta_3 \\[-10pt]
                    \\[8pt]
2\sim\left(
     \begin{array}{c}
       \frac{1}{\sqrt{2}}(\al_2\be_2-\al_3\be_3) \\
       \frac{1}{\sqrt{6}}(-2\al_1\be_1+\al_2\be_2+\al_3\be_3) \\
     \end{array}
   \right)\\[-10pt]
   \\[8pt]
3_1\sim\left(\begin{array}{c}
         \al_2\be_3+\alpha_3\beta_2 \\
         \al_1\be_3+\alpha_3\beta_1  \\
         \al_1\be_2+\alpha_2\beta_1  \\
        \end{array}\right)\\[-10pt]
        \\[8pt]
3_2\sim\left(\begin{array}{c}
         \al_3\be_2-\alpha_2\beta_3 \\
         \al_1\be_3-\alpha_3\beta_1  \\
         \al_2\be_1-\alpha_1\beta_2  \\
         \end{array}\right)
\end{array}\right.\\
&3_1\otimes3_2=&&1_2\oplus2\oplus3_1\oplus3_2&&\qquad
\text{with}\quad\left\{
\begin{array}{l}
1_2\sim\alpha_1\beta_1+\alpha_2\beta_2+\alpha_3\beta_3\\[-10pt]
        \\[8pt]
2\sim\left(
     \begin{array}{c}
       \frac{1}{\sqrt{6}}(2\al_1\be_1-\al_2\be_2-\al_3\be_3) \\
       \frac{1}{\sqrt{2}}(\al_2\be_2-\al_3\be_3) \\
     \end{array}
   \right)\\[-10pt]
        \\[8pt]
3_1\sim\left(\begin{array}{c}
         \al_3\be_2-\alpha_2\beta_3 \\
         \al_1\be_3-\alpha_3\beta_1  \\
         \al_2\be_1-\alpha_1\beta_2  \\
    \end{array}\right)\\[-10pt]
        \\[8pt]
3_2\sim\left(\begin{array}{c}
         \al_2\be_3+\alpha_3\beta_2 \\
         \al_1\be_3+\alpha_3\beta_1  \\
         \al_1\be_2+\alpha_2\beta_1  \\
    \end{array}\right)\\
\end{array}\right.
\end{aligned}
\eeq

\vspace{.5cm}
We will indicate this basis as the $D$-basis since in this basis both triplets and doublets contract diagonally to give the singlet according to the multiplication rules shown above. Other two $S_4$ basis are particularly important in flavour model building  and we will refer to them as the TB-basis and the BM-basis respectively. TB-basis is characterized by having the $T$ generator diagonal in the triplet and doublet representation.  All the $S_4$ elements in the TB-basis are obtained by the $D$-basis through a rotation
\beq
\label{TBOm}
X_{TB}=V \cdot X_D \cdot V^\dagger\,,
\eeq
where $X_{TB,D}$ are the $S_4$ element in the two different basis and $V=V_{\pi/4},U_\omega$  stays for transformations for the doublet and  triplet  representation, respectively.   They are given by
\beq
\label{Drot}
V_{\pi/4}=\frac{1}{\sqrt{2}}\left( \begin{array}{cc}  i&-i\\1&1\end{array}\right)\,,\qquad\qquad
U_\omega=\frac{1}{\sqrt{3}}\left( \begin{array}{ccc} 1&1&1\\1&\omega&\omega^2\\1&\omega^2&\omega\end{array}\right),
\eeq
with $\omega =e^{2 \pi i/3}$.

The BM-basis is characterized by having the $S$ generator diagonal in all the representations and is obtained by the $D$-basis through a rotation  equivalent to eq.~(\ref{TBOm}):
\beq
\label{BMOm}
X_{BM}=V \cdot X_D \cdot V^\dagger\,,
\eeq
where now $V=V_{\pi},U_{\pi/4}$ with
\beq
V_{\pi}=-\left( \begin{array}{cc}  1&0\\0&1\end{array}\right)\,,\qquad\qquad
U_{\pi/4}=\frac{1}{\sqrt{2}}\left( \begin{array}{ccc} \sqrt{2}&0&0\\0&i&i \\0&-1&1\end{array}\right)\,.
\eeq

In the following sections we will see that there is a tight correspondence between the $S_4$ basis chosen and the  flavour model  building realization. Even if physics is independent by the chosen basis, a framework may be easier implemented with a basis instead of with another.

All the models based on $S_4$ may be distinguished into two major classes: those ones that predict at LO a FD mass matrix and thus a lepton mixing matrix that is given by pure numbers, and those ones that do not. In the first class we recognize three subgroups: 
$i)$ models in which the lepton mixing arise by the correct combination between neutrino and charged lepton mixing matrix, $ii)$ models in which the TB scheme arises only by the neutrino sector, $iii)$ models in which the BM pattern arises only by the neutrino sector.  Typically models belonging to subclass $i)$ are realized within the $D$-basis, while models belonging to subclass $ii)$ within the $TB$-basis and those of subclass $iii)$ within the $BM$-basis. There is an additional class of models that can be thought between these two and it is characterized by models that predict the TM mixing pattern. Strictly speaking this pattern is a sort of partial FD mixing because only one eigenvector--and correspondingly one mixing angle--is fixed by the flavour symmetry.

Before describing more in details the models based on the $S_4$ discrete symmetry, we comment on the claim \cite{Lam:2008sh} that, in order to obtain the TB mixing ``without fine-tuning'', the finite group must be $S_4$ or a larger group containing $S_4$. For us this claim is not well grounded being based on an abstract mathematical criterium for a natural model (see also Ref.~\cite{Grimus:2009pg}). For us a model is natural if the interesting results are obtained from the most general lagrangian compatible with the stated symmetry and the specified representation content for the flavons. For example, in Ref.~\cite{Altarelli:2005yp,Altarelli:2005yx}, a natural (in our sense) model for the TB mixing is built with $A_4$ (which is a subgroup of $S_4$) by simply not including symmetry breaking flavons transforming like the $1'$ and the $1''$ representations of $A_4$. This limitation on the transformation properties of the flavons is not allowed by the rules specified in Ref. \cite{Lam:2008sh}, which demands that the symmetry breaking is induced by all possible kinds of flavons (note that, according to this criterium, the SM of electroweak interactions would not be natural because only Higgs doublets are introduced!). Rather, for naturalness we also require that additional physical properties like the VEV alignment or the hierarchy of charged lepton masses also follow from the assumed symmetry and are not obtained by fine-tuning parameters: for this actually $A_4$ can be more effective than $S_4$ because it possesses three different singlet representations 1, $1'$ and $1''$.

\boldmath
\section{$S_4$-Based Neutrino Models}
\label{sec:gen}
\unboldmath
In this core section we give an up to date overview of the $S_4$ models present in the literature according to the classification we did at the end of Sec.~\ref{sec:group}, between models that predict an exact FD lepton mixing at LO, and those that do not. We already said that in  the first group  there are the TB and BM mixing models, while the TM mixing one is in the middle and  we include it into the second group for reason that will be clear in the next sections.

\subsection{FD Models}

Here we introduce the theoretical motivation behind the  FD mixing patterns as already discussed in \cite{Lam:2007qc,Lam:2008sh,Altarelli:2009gn,Toorop:2011jn,deAdelhartToorop:2011re}.  The same arguments apply not only to $S_4$ but to any non-Abelian discrete symmetry. For this reason we will be as generic as possible following \cite{Toorop:2011jn,deAdelhartToorop:2011re}.

Let us consider a non-Abelian discrete group $G_f$, of elements $\{g_i\}$ with $i=1,...,d_G$ , being $d_G$ the dimension of the group. Let  $r$ be an  irreducible representation of dimension $n$. The $n\times n$ matrices  $r(g_i)$ will be the representations of the elements of $G_f$.  Suppose that $G_f$ has a given set of subgroups $G_{f_l}$, $l=1,..,h$ of dimensions $d_{G_l}$. Then each subgroup is generated by a set of $d_{G_l}$ $\{ g_{l_k}\}$ subset  of $\{g_i\}$. A $n\times n$ matrix $M$ is invariant under the subgroup $G_{f_l}$ if for each $g_{l_k}$, $k=1,..,d_{G_l}$,  we have that 
\beq
\label{rsym}
r(g_{l_k})^T \cdot M \cdot  r(g_{l_k})= M\,,
\eeq
if $M$ is symmetric, or 
\beq
\label{rher}
r(g_{l_k})^\dag \cdot M \cdot  r(g_{l_k})= M\,,
\eeq
if $M$ is hermitian. In general the irreducible representation $r$ of $G_f$ is  a reducible representation of the subgroup $G_{f_l}$ and  decomposes into $r_{l_1}\oplus \dots \oplus r_{l_m}$  irreducible  representations of  $G_{f_l}$. This means that exists a unitary transformation $\Omega_l$ that rotates  each  $r(g_{l_k})$ into a block diagonalizable form corresponding to $r_{l_1}(g_{l_k})\oplus \dots \oplus r_{l_m}(g_{l_k})$, or in other words that
\beq
\Omega_l^\dag \cdot r(g_{l_k})\cdot \Omega_l= \hat{r}(g_{l_k})\,,
\eeq
where we have indicated $\hat{r}(g_{l_k})=r_{l_1}(g_{l_k})\oplus \dots \oplus r_{l_m}(g_{l_k})$.  Thus eq.~(\ref{rsym}) becomes
\beq
\label{rsym}
  \hat{r}(g_{l_k})^T \cdot \hat{M} \cdot \hat{r}(g_{l_k})= \hat{M}\,,
\eeq
where we have indicated 
\beq
\hat{M}= \Omega_l^T \cdot  M \cdot  \Omega_l\,.
\eeq
Considering eq.~(\ref{rher}) we straightforwardly get
\beq
\hat{r}(g_{l_k})^T \cdot \hat{M} \cdot \hat{r}(g_{l_k})= \hat{M}\,,\qquad\qquad
\hat{M}= \Omega_l^\dag \cdot  M \cdot  \Omega_l\,.
\eeq
$\hat{M}$ maybe interpreted as the matrix $M$ in the $G_{f_l}$ basis,  being $G_{f_l}$  the subgroup of $G_f$ under which it is invariant.

Let us consider now the basis in which both neutrino and charged lepton mass matrices are diagonal. Since no degeneracy is present in any of the two sectors, if there is any residual flavour symmetry this has to be Abelian. Assuming that light neutrinos are Majorana particles such Abelian symmetry is limited to be $Z_2\times Z_2$. The reason is very simple:  the  Majorana nature of  neutrinos allows them to be charged only under a $Z_2$, giving them and odd or even flavour parity. However with only one $Z_2$, it is not possible to have a diagonal neutrino mass matrix --we are in the mass eigenstate basis-- and therefore it is necessary to introduce a further $Z_2$ to forbid off-diagonal entries. For what concerns the charged leptons, being Dirac particles, there is a infinite class of Abelian symmetries, both continuos and discrete, that may be chosen to get a diagonal charged lepton mass matrix. The minimal choices are obviously $Z_2\times Z_2$, $Z_2\times Z_2\times Z_2$ and $Z_3$. 

Coming back to the general discussion, it is clear that $G_f$ has to be read as the main flavour group, while the subgroup $G_{f_l}$  are only its  Abelian subgroups that define the flavour structure of neutrinos and charged leptons. As a consequence the representation $r_{l_i}(g_{l_k})$ are all one dimensional and consequently both $\hat{r}(g_{l_k})$ and $\hat{M}$ are diagonal. Finally $\Omega_l$ corresponds to  the matrix that diagonalizes $M$. In other words if $M$ is the neutrino (charged lepton) mass matrix $M_\nu$  $ (M^\dag_e M_e)$ invariant under $G_{f_\nu} (G_{f_e})$ subgroup of $G_f$, the lepton mixing is given by
\beq
U_{lep}=\Omega_e^\dag \cdot \Omega_\nu\,.
\eeq
We automatically get the important information that $G_e$ cannot coincide with $G_\nu$ because otherwise $U_{lep}$ would coincide with the identity matrix.

 Any FD lepton mixing is obtained by choosing a non-Abelian discrete $G_f$ and consequently its subgroups  $G_\nu$ and $G_e$ for the neutrino and charged lepton sector respectively, that fixes unambiguously $\Omega_\nu$ and $\Omega_e$. 
 
 Finally a remark: when we say that we fix a subgroup $G_{f_l}$ of $G_f$ generated by the subset $\{ g_{l_k}\}$ we mean up to subgroups generated by  elements belonging to the same classes of $\{ g_{l_k}\}$. Since all the elements belonging to the same class may be obtained  by acting on only one element of that class, it turns out that the subgroups originated by elements of the same class give rise to a series of  $\Omega_l$ that are identical up to permutation of raws and columns.
 
 Let us now consider the case of $S_4$: by looking at Tab.~\ref{tabcht} and according to what said so far, the residual symmetry $G_\nu\sim Z_2\times Z_2$  in the neutrino sector has to arise by one element of $\mathcal{C}_2$ and one element of $\mathcal{C}_4$. If we take $S^2$ for the first and $T S T$ for the latter and impose that $M_\nu$ is invariant under both of them, according to eq.~(\ref{rsym}) we get
 \beq
 \label{massnu}
 M_\nu =\left(\begin{array}{ccc}a&0&0\\ 0& b&c\\ 0&c&b\end{array} \right)\,.
 \eeq 
 This matrix is diagonalized by a maximal rotation in the sector $23$ by means of  $R_{23}(\pi/4)$ using the notation introduced in eq.~(\ref{parmix}). According to what we have said, taking other two elements of the same classes we expect the same structure of eq.~(\ref{massnu})  but involving different columns/raws: indeed if we consider $T S^2 T^2$ of $\mathcal{C}_2$ and $S T S^2$ of $\mathcal{C}_4$, $M_\nu$ is diagonalized by $R_{13}(\pi/4)$, while, taking $S^2 TS^2 T^2$ and $S T^2$, the rotation is  $R_{12}(\pi/4)$. Finally all the other combination lead to the same structures or to the limiting cases for which the off-diagonal entries are put to zero, that is  $c=0$ in eq.~(\ref{massnu}).
 
 For what concerns the  charged lepton sector  we have now two options: we may take the subgroup generated by an element of the $\mathcal{C}_3$, corresponding to $G_e\sim Z_3$, or  by an element of $\mathcal{C}_5$ corresponding to $G_e\sim Z_4$.  By taking $T$ of  $\mathcal{C}_3$ and requiring that $M_e M_e^\dag$ is invariant under $T$, where $M_e$ is the charged lepton mass matrix in the LR notation, forces
\beq
M_e M_e^\dag  =\left(\begin{array}{ccc}f&g&g^*\\ g^*& f&g\\ g&g^*&f\end{array} \right)\,,
\eeq 
that translates into
\beq
M_e =\left(\begin{array}{ccc}h_0&h_1&h_2\\ h_2 &h_0&h_1\\ h_1&h_2&h_0\end{array} \right)\cdot V_R\,,
\eeq 
being $V_R$ an arbitrary non-physical rotation in the right handed charged sector\footnote{Notice that $V_R$ is non-physical as long as we deal with the SM. In a supersymmetric context $V_R$ may be related to the corresponding rotation in the right handed charged slepton sector with no trivial consequences for the phenomenology of the model.}.
The latter is diagonalized by the matrix $U_\omega$ introduced  in Sec.~\ref{sec:group} to switch basis  from the D-basis to the TB-basis and it  has been defined in eq.~(\ref{Drot}).  It is clearly not surprising that in this case by combining $U_\omega$ with $R_{23}(\pi/4)$ according to
\beq
U_{lep}= U_\omega^\dag\cdot R_{23}(\pi/4)\,,
\eeq  
the final lepton mixing is the TB one. Notice that in the TB-basis the $T$  generator element  of $\mathcal{C}_3$ is diagonal, thus $M_e$ is diagonal and all the mixing pattern arises from the neutrino sector. From this the name of the basis.
   
Let us now consider the case in which  $G_\nu$  is generated by  the elements $S^2 TS^2 T^2$ of $\mathcal{C}_2$ and $S T^2$  of $\mathcal{C}_4$. We already said that this choice leads the following neutrino mass matrix
\beq
\label{massnu2}
M_\nu =\left(\begin{array}{ccc}a&b&0\\ b& a&0\\ 0&0&c\end{array} \right)\,,
\eeq 
diagonalized by $\Omega_\nu\sim R_{12}(\pi/4)$. In this case, for the charged leptons, we can take $G_e \sim Z_4$: this means that we require $M_e M_e^\dag$ to be invariant under, for example, the action of  $S$, element of $\mathcal{C}_5$. This leads to
\beq
\label{mchq}
M_e M_e^\dag  =\left(\begin{array}{ccc}f &0&0\\ 0& g& i h\\ 0&-i h&g\end{array} \right)\,,
\eeq 
where $f$, $g$ and $h$ are real parameters. This matrix is diagonalized by a $R_{23}(-\pi/4)$ up to a phases: more precisely by:
\beq
\label{V23max}
\tilde{R}_{23}(-\pi/4)= \left(\begin{array}{ccc}1 &0&0 \\0& \frac{1}{\sqrt{2}}&- \frac{1}{\sqrt{2}}\\0& \frac{1}{\sqrt{2}}& \frac{1}{\sqrt{2}}\end{array} \right) \cdot \left(\begin{array}{ccc}1 &0&0 \\0& -i &0\\0& 0&1\end{array}\right) \,.
\eeq
In this way by combining  eq.~(\ref{V23max}) with $R_{12}(\pi/4)$ arising by the neutrino sector we get the BM pattern of eq.~(\ref{genBM}). The diagonal phase matrix may be absorbed by a  field redefinition.
 
Let us concentrate  on the consequence of eq.~(\ref{mchq}) for $M_e$ once we have absorbed the phases of  eq.~(\ref{V23max}). $M_e$ presents  the generic structure
\beq
M_e =\left(\begin{array}{ccc} g_0 e^{i \varphi_g}&0&0\\ 0 &h_0 e^{i\varphi_0}&h_2  e^{i\varphi_2} \\ 0& h_1e ^{i (m \pi+\varphi_1)}& \sqrt{h_2^2+h_0^2-h_1^2}e ^{i (n\pi+\varphi_2)}\end{array} \right)\cdot  V_R\,,
\eeq 
where $g_0, h_0,h_1,h_2$ are real parameters, $\varphi_{g,i}$ their phases,  $m,n$ integer numbers and $V_R$ an arbitrary rotation in the right handed sector. Here we have considered the case $h_2^2+h_0^2-h_1^2\geq0$. However the case  $h_2^2+h_0^2-h_1^2<0$  is easily got by changing the sign and replacing $n$ with $n+1$. $M_e$ describes the charged lepton mass matrix, so we need 3 different eigenvalues.  For the most generic case $M_e M_e^\dag$ has three distint eigenvalues
\beq
\begin{aligned}
m^2_1 &= g_0^2\,,\\
 m^2_{2}&= h_2^2 +h_0^2 -\sqrt{ h_2^4+  h_0^2 h_1^2+ h_2^2( h_0^2- h_1^2)+ 2 h_2 h_1 h_0 \sqrt{h_2^2+h_0^2-h_1^2}\cos(m-n)\pi}\,,\\
  m^2_{3}&= h_2^2 +h_0^2 +\sqrt{ h_2^4+  h_0^2 h_1^2+ h_2^2( h_0^2- h_1^2)+ 2 h_2 h_1 h_0 \sqrt{h_2^2+h_0^2-h_1^2}\cos(m-n)\pi}\,.
\end{aligned} 
\eeq
However in many models, it is easier getting high symmetric mass matrices:  consider for example the case in which $h_0= h_1$. In this case the two last eigenvalues becomes
\beq
m^2_{2,3}= h_2^2 +h_0^2- (\pm)\sqrt{h_2^4+h_0^4 + 2 h_2^2 h_0^2 \cos(m-n)\pi}\,.
\eeq 
In order to have $m_2^2\neq0$ we need $m-n=1+ 2 k \pi$, with $k$ an integer. Thus taking $m=1$, $n=0$ and $k=0$ --and $V_R=1$-- leads us to the simplest form for $M_e$ we may get in the D-basis
\beq
M_e =\left(\begin{array}{ccc} \tilde{g}_0 &0&0\\ 0 &\tilde{h}_0 &\tilde{h}_2   \\ 0& -\tilde{h}_0  & \tilde{h}_2  \end{array} \right)\,,
\eeq 
where now $\tilde{g}_0,\tilde{h}_i$ are complex.  This texture for the charged lepton mass matrix has been used in Ref.~\cite{Toorop:2010yh}. Notice that in that context it was deduced by simple phenomenological motivations.  On the contrary in  the $D$-basis the  most generic charged lepton mass matrix invariant under $Z_4$ is not  easily obtained in a concrete model. Once again the choice of the basis helps: in the BM-basis  $S$, generator of $Z_4$ is diagonal, thus $M_e$ is diagonal with 3 distinct mass eigenvalues and all the lepton mixing arises by the neutrino sector.
  
Recently, on the basis of only theoretical group theory considerations, the authors of Ref.~\cite{Hernandez:2012ra} showed that in the case of $S_4$ new patterns predicting $\theta_{13}\neq0$ may be obtained. This could seem in contrast with what we have showed so far and with what is stated in Refs.~\cite{Toorop:2011jn,deAdelhartToorop:2011re}. However, this is not the case. Indeed, under the assumption that neutrinos are Majorana particles, their mass matrix presents always a $Z_2\times Z_2$ symmetry. One of the two $Z_2$ could be accidental and not belonging to the flavour non-Abelian group. It has been shown in Ref.~\cite{Hernandez:2012ra}, from general group theoretical considerations, that certain correlations among the mixing angles follow from the residual symmetries of the charged lepton and neutrino sectors, once the initial flavour symmetry is broken. As a result, the relations showed so far and in Refs.~\cite{Toorop:2011jn,deAdelhartToorop:2011re} are special cases of the more general expressions recovered in Ref.~\cite{Hernandez:2012ra}. Although we find the results in Ref.~\cite{Hernandez:2012ra} interesting and potentially inspiring for new strategies in neutrino model building, the construction of a concrete model predicting the correlations showed in Ref.~\cite{Hernandez:2012ra} could be quite challenging: indeed, when only one of the two $Z_2$ symmetry factors of the neutrino mass matrix is contained in the flavour group $G_f$, it is necessary the accidental presence of a $Z_2$ factor aside the flavour group, $G_f\times Z_2$, and this is subjected to a well precise alignment in the flavour space of the flavon vevs and to a specific choice of the transformation properties of all the fields under $G_f$.

\subsection{Breaking the Flavour Symmetry}
\label{sec:SB}
  
From a group theory point of view, it is quite natural to assume that a given non-Abelian discrete group $G_f$ breaks down to two different subgroups $G_\nu$ and $G_e$ in the neutrino and charged lepton sector, respectively. However from a practical point of view it is not easy that a model accounts for the correct mechanism to break the symmetry.
 
 Typically the breaking of the flavour symmetry is felt by the SM fermions, with the addition of the  right handed neutrinos, through their couplings with scalar fields that transform not trivially under the flavour symmetry, usually addressed as flavons. In the most traditional scenarios the flavons are singlets of the SM gauge groups and bring only flavour charges, while the SM boson  Higgs field is not charged under the flavour group, or better  said it is not charged under the non-Abelian part of the flavour group, since it is often  charged under additional Abelian symmetries. Additional scalars necessary  to induce neutrino masses (like $SU(2)_L$ triplets) behave in general  as the Higgs field with respect to the flavour group. 
 
Consequently, these models are written in terms of non-renormalisable effective operators suppressed by suitable powers of the cut-off. Since in general the higher-dimensional operators cannot be neglected, it is necessary to take into account at least the NLO  corrections --sometimes even NNLO-- as long as an UV renormalizable version of the models is not studied.  In the most  \emph{traditional}  spirit we expect that the flavons develop VEVs inducing the spontaneous breaking of the  original flavour symmetry, thus giving rise to fermion mass textures. However one can think to ED scenarios in which the flavons leave in the bulk while the SM fermions on the boundaries. In this case adequate boundary conditions (BCs) could  break the symmetry and make the desired job. 
In this brief discussion we will not consider this latter case, but we acknowledge it as an alternative to the traditional dynamics of the flavour symmetry breaking, that in our opinion is one of the weakest point of these models.
 
Let us consider the spontaneous breaking of the flavour symmetry:  since $G_f$ needs to be broken down to $G_\nu$ and $G_e$ in the neutrino and charged lepton sector, respectively, and $G_\nu\neq G_e$, we need two distinct set of flavons, denoted as $\{\phi_\nu\}$ and $\{\phi_e\}$, that implement the breaking chains $G_f\to G_\nu$ and $G_f\to G_e$  and communicate the breaking of the symmetry to neutrinos and charged leptons. We used the notation $\{\phi_{\nu,e}\}$ to be as general as possible, meaning that for each breaking we may have more than one flavon field. If now, for concreteness, we consider a single flavon $\phi_l\sim r$, where $r$ is an  irreducible representation of $G_f$, denoting as $\langle \phi_l \rangle$ its VEV, we have that $G_f\to G_{f_l}$ if for any $r(g_{l_k}) $  it holds 
\beq
\langle \phi_l \rangle \cdot  r(g_{l_k})=  r(g_{l_k})\,,
\eeq
being $\{g_{l_k} \}$ the elements of $G_{f_l}$. For example, if we work in the $D$-basis  a triplet  $\phi$ of $S_4$  induces the breaking $S_4\to Z_2\times Z_2$ where $Z_2\times Z_2$ is generated by $S^2$ and $TST $ if  $\langle \phi_l \rangle$ is aligned as $(1,0,0)$. Similarly $\phi$ induces the breaking $S_4\to Z_3$ generated by $T$ if its VEV aligns as $(1,1,1)$.

Given a set of flavons, the spontaneous breaking of $G_f$ to \emph{one} of its subgroup is naturally realized: typically we need a negative mass square term in the flavon scalar potential  and then the relative size of the potential parameters determine the subgroup in which $G_f$ breaks. On the contrary, in the most general case, $G_f$ does not break to two \emph{distinct} subgroups: this turns into the necessity to separate the scalar potential of  $\{\phi_\nu\}$ from that of   $\{\phi_e\}$. The easiest way to realize this is by introducing additional Abelian symmetries that also prevent $\{\phi_\nu\}$ coupling to the charged lepton sector and $\{\phi_e\}$ coupling to the neutrino one. A further help is provided by promoting the models to be supersymmetric, even if it is not mandatory \cite{Bazzocchi:2008ej}: in the supersymmetric case by imposing the $U(1)_R$ continuous symmetry and by introducing an additional set of scalar fields, usually called driving fields, the desired breaking chains are relatively simply realized. 

The drawback of splitting up the potential is that it generates accidental continuous symmetries that once broken give rise to massless goldstone bosons. The problem of the massless particles is avoided by introducing soft flavour symmetry breaking terms, but the origin of such soft terms has not been addressed so far. 

Finally we remind that the scalar  potential may be splitted embedding the models in large EDs: in this way the two sets of flavons live one on the IR brane and the other on the UV one and the interference between the two sectors are suppressed by the size of the extra dimensions. Notice that this approach is completely different with respect to breaking the flavour symmetry by means of BCs: indeed in this case the EDs setting is used to separate the two set $\{\phi_\nu\}$ and $\{\phi_e\}$, making them living into different branes and the flavour symmetry breaking  is induced by flavon VEVs as usual.
 
In conclusion the correct flavour symmetry breaking necessary to induce  FD mixing patterns require additional ingredients that make the models more baroque and less appealing: extra  symmetries, extra dimensions, extra scalar fields, unexplained soft terms. This is a serious drawback for FD mixing models, because as much as natural they seem under group theoretical considerations as less natural they look after a concrete model building realization. For these reasons no FD mixing models represent a complementary and compelling approach to account for the flavour puzzle: the underlying flavour symmetry still maintain a certain level of predictivity but its breaking is easily realized.

 \subsection{General Criteria to Classify Flavour Models}
 \label{criteria}

In the next sections we will summarize the main features of the models  that  at LO predict TB  and  BM lepton mixing, respectively.
In order to discriminate among them and to identify the most complete ones, we list a series of general criteria typically addressed in flavour model building:
\begin{itemize}
\item[a)] \emph{realistic realization:} we have said that  flavour breaking patterns are not easily realized. In principle one could build a flavour model assuming unjustified  vacuum alignments hoping that it will exist a scalar potential or a mechanism to  explain them. Models that do not enter into the details of  flavour symmetry  breaking  are incomplete and weak because based on a hypothesis that could not be true;
\item[b)] \emph{predictions in the neutrino sector:} 
there is a plethora of way to explain the origin of neutrino masses that reflects in different phenomenologies. Moreover the flavour symmetry may be implemented in such a way to get interesting predictions for neutrino observables, such as $0\nu2\beta$ decay, the kind of hierarchy,  the reactor angle and CP-phases. Predictions are a \emph{plus value} for this kind of models;
\item[c)] \emph{GUT embedding:} 
it is quite appealing to try to combine flavour symmetry with grand unified theories (GUT). This is deeply challenging because in GUT models quarks and leptons are (partially) unified while quark and lepton mixing  and mass hierarchies are completely different. Models that give a unified description of quarks and leptons can be considered more complete. 
\item[d)] \emph{quark sector:} being less ambitious than at point $c)$, we may wonder if quark mixing and mass hierarchies may be described by the same flavour symmetry as leptons, even if not in a GUT context;
\item[e)] \emph{further predictions:} models satisfying criteria $c)$ or the simpler $d)$ may have interesting prediction in the quark sector (such as the Gatto-Sartori-Tonin (GST) \cite{Gatto:1968ss} or the Georgi-Jarlskog (GJ) \cite{Georgi:1979df} relations) or interesting correlations between charged lepton and down quark masses. Moreover fermion masses may satisfy sum rules.
\end{itemize}

\begin{table}[!h]
\vchcaption{Model properties according to the criteria enunciated in Sec.~\ref{criteria}. FD models are grouped into two classes,  namely { \bf TB}  and {\bf BM}, referring to those predicting TB  and BM patterns as the LO lepton mixing matrix, respectively. In a  third group, {\bf no FD}, we put all the $S_4$ flavour models that do not have at LO a FD lepton mixing matrix. ``$V(\phi,h)$'' stays for the flavour and Higgs scalar potential study; ``$\nu$'' indicates the neutrino sector properties; ``GUT'' if the model has a GUT framework; ``$q$'' if it describes also quark sector; ``$q_{rel}$'' if it predicts quarks relations such as GST and/or GJ; finally  under ``extra'' we grouped all the other relevant features. If the latters are absent we put a  ``-''. For what concerns neutrino porperties  ``type-X'' stays for the kind of See-Saw, ``ISS'' for Inverse See-Saw and ``W.op.''  for Weinberg operator. }
\label{table:models}
\begin{tabular}{@{}ccccccc@{}}
\hline
  &&&&&& \\[-3mm]
  												&  $V(\phi,h)$		& $\nu$ 						& GUT 			& $q$	& $q_{rel}$	& extra\\[1mm]
  \hline
  \hline
  &&&&&&  \\[-3mm]
  {\bf TB}\\[1mm]
  \hline
  &&&&&& \\
  \cite{Koide:2007sr} 					& roughly				& type I								& no				& no		& no				& $SU(3)$ embedding  \\
  \cite{Bazzocchi:2008ej} 			& yes					& type I+II						& no 				& no 		& no				& no susy \\
  \cite{Ishimori:2008fi}				& no						& type I							& $SU(5)$		& yes	& no				& complicated scalar sector \\
  \cite{Bazzocchi:2009pv}	+ 	\cite{Bazzocchi:2009da}	& yes					& type I+II+III+W.op.						& no				& yes	& no				& -\\
  \cite{Ding:2009iy}					& yes					& type I						& no				& yes	& no				& -\\
  \cite{Dutta:2009bj}					& no						& type II						& $SO(10)$	& yes	& no				& -\\
  \cite{Meloni:2009cz}				& yes					& type I							& no				& no		& no				& -\\
  \cite{Adulpravitchai:2010na}	& BCs in ED			&				type I					& $SO(10)$	& yes	& no				& 8D,orbifolding\\
  \cite{Hagedorn:2010th}			& yes					& type I							& $SU(5)$		& yes	& yes			& -\\
  \cite{Ishimori:2010xk}				& yes					& type I							& $SU(5)$		& yes	& no				& only Cabibbo in $V_{CKM}$\\
  \cite{Ding:2010pc}					& yes					& type I+III					&$SU(5)$		& yes	& no				& -\\
  \cite{Zhao:2011pv}					& yes					& type I, NH, $m_1=0$	&no				& no		& no				& -\\
  &&&&&&\\
  \hline
  \hline
   &&&&&& \\[-3mm]
  {\bf BM}\\[1mm]
  \hline
  &&&&&& \\
  \cite{Altarelli:2009gn}				& yes					& lower bound for $m_1$& no			& no		& no				& -\\
  \cite{Toorop:2010yh}				& yes					& type II						& PS				& yes	& yes			& QLC\\
  \cite{Patel:2010hr}					& no						& type II						& $SO(10)$	& yes	& no				& QLC, renormalizable\\
  \cite{Meloni:2011fx}					& yes					& W.op.				& $SU(5)$ 	& yes	& no				& $b-\tau$, 5D orbifolding\\
    &&&&&&\\
  \hline
  \hline
   &&&&&& \\[-3mm]
    {\bf No FD}\\[1mm]
  \hline
  &&&&&& \\
 \cite{Hagedorn:2006ug}  			& yes					& type I							& no				& yes	& no				& embed. in $SO(10)\times G_f$  \\ 
   \cite{Morisi:2010rk} 				& yes					& W.op., $\theta_{atm}=\pi/4$&no&yes	&no				& $\ltimes$ product of many $Z_2$\\
   \cite{King:2011zj}					& yes					& type I, TM					&no				&no		&no				& new sum rule\\
      \cite{Dorame:2012zv}					& roughly					& ISS					&no				&yes		&no				& new sum rule\\
    &&&&&&\\
  \hline
\end{tabular}
\end{table}

\subsubsection{ TB Models}
\label{TBmodel}

Here we deal with the models that predict TB mixing at LO order. Armed by the criteria exposed in Sec.~\ref{criteria} and by Tab.\ref{table:models} it is easier to describe the different features and properties.

In all the models the flavour symmetry is represented by a product of $S_4$ and additional Abelian symmetries: only in the model in Ref.~\cite{Ishimori:2010xk}, a Frogatt-Nielsen continuous symmetry is introduced, while in all the others discrete symmetries have been used. Moreover, all of them, a part from the realisation in Ref.~\cite{Bazzocchi:2008ej}, are supersymmetric.  As a result, we did not list these features in Tab.\ref{table:models}.

The models in Refs.~\cite{Ishimori:2008fi} and \cite{Dutta:2009bj} do not satisfy criterium $a)$, while in Ref.~\cite{Koide:2007sr} only a rough analysis has been done. In this paper an embedding in $SU(3)$ is necessary to get the so-called Koide relation \cite{Koide:1982si,Koide:1982ax,Koide:1983qe} among the charged lepton masses, 
\beq
\dfrac{m_e+m_\mu+m_\tau}{\left(\sqrt m_e+\sqrt m_\mu+\sqrt m_\tau\right)^2}=\dfrac{2}{3}\,,
\eeq
that is very well verified for pole masses. However, the VEV alignment is \emph{ad hoc}.  In the model in Ref.~\cite{Ishimori:2008fi}  neither the flavour nor the GUT scalar potentials are studied and the scalar sector to get fermion charged masses is deeply cumbersome. Similar situation for the model in Ref.~\cite{Dutta:2009bj}: the scalar potentials are not studied and neutrino masses are assumed to be generated by the type-II  See-Saw mechanism. On the contrary, for all the other models the scalar potential is analyzed in details. In particular, in the model in Ref.~\cite{Hagedorn:2010th}, a really exhaustive study of both the scalar potentials is performed, both at LO and NLO (in Ref.~\cite{Antusch:2011sx} a variation of the model in Ref.~\cite{Hagedorn:2010th} has been presented with the focus on the CP violating Dirac phase). In  Ref.~\cite{Adulpravitchai:2010na}, $S_4$ is obtained by orbifolding a 8-dimensional space  and then the  flavour symmetry is broken by BCs.

For what concerns point $b)$, in the majority of the models, neutrino masses are induced by type I See-Saw mechanism. However, in Ref.~\cite{Dutta:2009bj}, the type II mechanism is implemented, while in Ref.~\cite{Ding:2010pc}, there is an interplay between type-I and type-III. In Ref.~\cite{Bazzocchi:2008ej}, neutrino masses  are induced by the interplay between type-I and II See-Saw.  Refs.  \cite{Bazzocchi:2009pv}	-\cite{Bazzocchi:2009da} collect different scenario those more mechanisms to generate neutrino masses are involved.
All the models have NLO corrections that may give a $\theta_{13}$ in agreement with the most recent data. However, NLO corrections are under control only in those models that satisfy criteria $a)$. Quite interestingly in Ref.~\cite{Zhao:2011pv}, only the NH case with $m_1=0$ is predicted.

All the models in Refs.~\cite{Ishimori:2008fi,Hagedorn:2010th,Ishimori:2010xk,Ding:2010pc,Dutta:2009bj,Adulpravitchai:2010na} are embedded in a GUT scenario: the first four in $SU(5)$, while the last two in $SO(10)$. All the models fit quark masses, even if the model in Ref.~\cite{Ishimori:2008fi} has a complicated set of scalars to fit the charged masses. For what concerns the mixings, the model in Ref.~\cite{Ishimori:2010xk} predicts only the Cabibbo angle, while the other models fit all the CKM matrix.  However, only the model in Ref.~\cite{Adulpravitchai:2010na} satisfies the GST and GJ relations, thanks to a specific set of messengers.

\subsubsection{BM Models}
\label{BMmodel}

The number of models  that at LO predicts exact BM mixing in the context of $S_4$ is relatively small. All the models are embedded in a supersymmetric scenario and have additional Abelian symmetries: we did not put these features in Tab. \ref{table:models}, as we did not for the TB mixing models. As done in Sec.~\ref{TBmodel} we summarize them by means of the \emph{criteria}  given in Sec.~\ref{criteria}.

In Ref.~\cite{Patel:2010hr}, a renormalizable model with the addition of messenger fields is built. However, both the flavon and the Higgs scalar potential are not studied. On the other hand, all the other models in this category satisfy criterium $a)$.

For what concerns the neutrino sector, in the model in Ref.~\cite{Altarelli:2009gn} a lower bound for the lightest neutrino is recovered, in the context of the type I See-Saw. In the models in Refs.~\cite{Toorop:2010yh}  and  \cite{Patel:2010hr}, the dominant contribution to neutrino masses arises from the type-II See-Saw: both the models are embedded into a GUT context, Pati-Salam and $SO(10)$, respectively, and therefore in general both type-I and type-II See-Saw contributions are present; the dominance of type-II must be suitably justified. However, only the model in Ref.~\cite{Toorop:2010yh} studies in details the Higgs scalar potential to check the validity of the type-II dominance, while it is simply assumed in Ref.~\cite{Patel:2010hr}. Finally, in Ref.~\cite{Meloni:2011fx}, neutrino masses can be induced by either the Weinberg operator or the type-I See-Saw.

In addition to the models in Refs.~\cite{Toorop:2010yh}  and  \cite{Patel:2010hr}, also that one in Ref.~\cite{Meloni:2011fx} is embedded in a GUT scenario, that in this case is $SU(5)$. Moreover this model presents an extra dimensional setup with the fifth dimension compactified on an orbifold, mainly to provide an explanation to the doublet-triplet splitting problem.

The non-GUT model, in Ref.~\cite{Altarelli:2009gn}, does not deal with quarks, while for the other models, the quark sector is automatically accounted for by the GUT embedding. In particular, the model in Ref.~\cite{Toorop:2010yh} presents at LO the GJ relation at the GUT scale, slightly broken once NLO effects are introduced. In Ref.~\cite{Patel:2010hr}, the GJ relation is reproduced only approximately even at LO,  since $m_\mu=-3 m_s$ is obtained assuming the smallness of a contribution. In Ref.~\cite{Meloni:2011fx}, only the $b-\tau$ unification is obtained at the GUT scale but it is preserved even at NLO. Finally in all the models, a part for that one in Ref.~\cite{Patel:2010hr}, a Froggatt-Nielsen $U(1)$ symmetry is introduced to explain the charged fermion mass hierarchies.

\subsubsection{No FD Models}

As already stated in Sec.~\ref{sec:SB}, flavour models based on a discrete  non-Abelian  flavour symmetry may maintain a good level of predictivity even if they do not reproduce a FD lepton mixing at LO. This is due to the reduced number of free parameters dictated by the flavour invariants. All the criteria in Sec.~\ref{criteria} may be extended even at these models, even if  point $a)$ is clearly more easily realized. 

In Ref.~\cite{Hagedorn:2006ug}, it is presented a model embeddable in a GUT $SO(10)\times G_f$, being $G_f\sim SO(3)_f$ or $SU(3)_f$ continuous flavour symmetry. The low energy Higgs doublets transform not trivially under $S_4$, thus the flavon and the Higgs sector need to be treated together and the $S_4$ breaking scale coincides with the EW one. Two numerical examples are provided for which the mass matrix textures lead to realistic lepton and quark masses and mixings. Furthermore, it is discussed the Higgs potential that allows for the appropriate VEV configurations required for such textures.

The model in  Ref.~\cite{Morisi:2010rk}  is a supersymmetric model in which the flavon and the Higgs sectors are treated together: a large set of $SU(2)$ Higgs doublets transforming non-trivially under the flavour symmetries is present. Indeed the model have six Higgs doublets and the minimization of the potential gives a residual $\mu-\tau$ symmetry that predicts maximal atmospheric angle and vanishing reactor angle. Both quarks and leptons mass hierarchies and mixings are fitted, and neutrino masses are induced by mean of the Weinberg operator. However, the full flavour symmetry of the model is baroque, being $G_F = S_4 \times Z_{3q} \times Z_{2q} \rtimes (Z_{2e} \times Z_{2\mu} \times Z_{2\tau} )$ and represents a weak point of the model. 

In Ref.~\cite{King:2011zj}, it is described an  $S_4$  supersymmetric model where the scalar VEVs break $S_4\to Z_2$  in the neutrino sector and to $Z_3$ in the charged lepton one. In this way, when charged lepton are diagonal, an exact TM lepton mixing is obtained and the mixing angles are expressed in term of deviations from the TB ones, namely $s$, $a$ and $r$ for the solar, atmospheric and reactor angle respectively. The interesting smoking-gun signature of such scenario is $s \simeq 0$ and  the sum rule $ 2a + r \cos\delta \simeq 0$, being $\delta$ the CP-Dirac phase.

Finally  Ref.~\cite{Dorame:2012zv}  presents  a supersymmetric model that justifies from first principles an interesting  new sum rule for neutrino masses in the context of the inverse See-Saw (ISS) mechanism. Once more the drawback of the model is a very large flavon sector and consequently a really complicated scalar flavon potential.

\section{Phenomenology}
\label{sec:Pheno}

In this section we deal with the phenomenological signatures of the models described in the previous sections. Without entering into details of a specific model, we present a general analysis that could be used as a guide to perform a more detailed study. We mainly concentrate on the presence of specific sum rules for the neutrino masses, on some neutrino observables, such as the sum of the neutrino masses, the kinematic electron neutrino mass in the single beta decay and the neutrino-less double beta decay effective mass, and on some lepton flavour violating transitions, like $\mu\to e\gamma$. These observables could be very useful to test the validity of flavour models and to distinguish one model from the other.

\boldmath
\subsection{Sum Rules and Related Observables}
\unboldmath

In the context of FD mixing pattern, the neutrino mass matrices depend only on three complex parameters (cfr. with eqs.~(\ref{MMTB}) and (\ref{MMBM})): three real parameters describe the neutrino masses and the remaining two corresponding to the Majorana phases. Indeed, the three mixing angles are fixed and independent on the parameters of the mass matrices and the Dirac CP phase is undetermined, being $\theta_{13}=0$.

Often, in flavour models based on non-Abelian discrete symmetries, only two independent complex parameters enter the neutrino mass matrix and therefore the three neutrino masses undergo to a well-defined correlation. Up to now, four different sum rules have been presented in literature \cite{Barry:2010yk,Dorame:2011eb}:
\beq
\begin{aligned}
m_1								&=\alpha\,m_2+\beta\,m_3\\
\dfrac{1}{m_1}				&=\alpha\,\dfrac{1}{m_2}+\beta\,\dfrac{1}{m_3}\\
\sqrt{m_1}						&=\alpha\,\sqrt{m_2}+\beta\,\sqrt{m_3}\\
\dfrac{1}{\sqrt{m_1}}		&=\alpha\,\dfrac{1}{\sqrt{m_2}}+\beta\,\dfrac{1}{\sqrt{m_3}}\,,
\end{aligned}
\eeq
where $m_i$ are the neutrino mass eigenvalues and $\alpha$ and $\beta$ two real and positive parameters, depending on the details of a specific model.

The first of these relations is usually present in model where the neutrino masses are described by the Weinberg Operator  or in the context of the type II See-Saw mechanism (see for example Refs.~\cite{Bazzocchi:2009pv,Bazzocchi:2009da}); the second and third sum rules are instead described in model with type I and type III See-Saw mechanism (see for example Refs.~\cite{Bazzocchi:2009da,Ding:2009iy}); while the last relation arises in the context of the inverse See-Saw (see for example Ref.~\cite{Dorame:2012zv}).

Once a specific mass ordering is chosen, it is possible to express the heaviest states in terms of the lightest ones and the mass squared differences: in the case of the normal ordering (NO), one can write
\begin{equation}
m_2=\sqrt{m_1^2+\Delta\,m^2_{sol}}\,,\qquad\qquad
m_3=\sqrt{m_1^2+\Delta\,m^2_{atm}}
\label{massesNO}
\end{equation}
while in the case of inverse ordering (IO), they are
\begin{equation}
m_1=\sqrt{m_3^2+\Delta\,m^2_{atm}}\,,\qquad\qquad
m_2=\sqrt{m_3^2+\Delta\,m^2_{sol}+\Delta\,m^2_{atm}}\,.
\label{massesIO}
\end{equation}
As a result, entering these expressions into the sum rule of a specific model, all the three neutrino mass eigenvalues are determined.

The precise determination of the three neutrino masses at experiments would represent a strong constraint for neutrino flavour models, testing for each model the corresponding sum rule. However, the present (and the expected future) sensibility is too low to allow such test. On the other hand, the presence of these correlations among the neutrino masses have an interesting impact on other observables: the sum of the absolute neutrino masses $\Sigma$, the kinematic electron neutrino mass in beta decay $m_\beta$ and the neutrino-less double beta decay effective mass $\mean{m_{ee}}$:
\beq
\begin{aligned}
\Sigma\,						&=\,\sum_{k=1}^{3}\,m_k\,,\\
m_{\beta}\,					&=\,\sqrt{\sum_{k=1}^3\,\left|U_{ek}\right|^2\,m^2_k}\,,\\
\langle m_{ee}\rangle 	&=
\left|\sum_k U^2_{ek}\,m_k\right|
= \left|c^2_{12}\, c^2_{13}\, m_1+
s^2_{12}\, c^2_{13}\, e^{i\alpha_{21}}\,m_2+
s^2_{13}\, e^{i\alpha_{31}}\,m_3\right|\,,
\end{aligned}
\eeq
where $U_{ek}$ are the elements of the first row of the lepton mixing matrix, $c_{ij}$ and $s_{ij}$ refer to cosines and sines of $\theta_{ij}$, while $\alpha_{21,31}$ are the Majorana phases in the usual convention.

\begin{figure}[h!]
\centering
\subfigure[$0\nu2\beta$ effective mass.]
{\includegraphics[width=7cm]{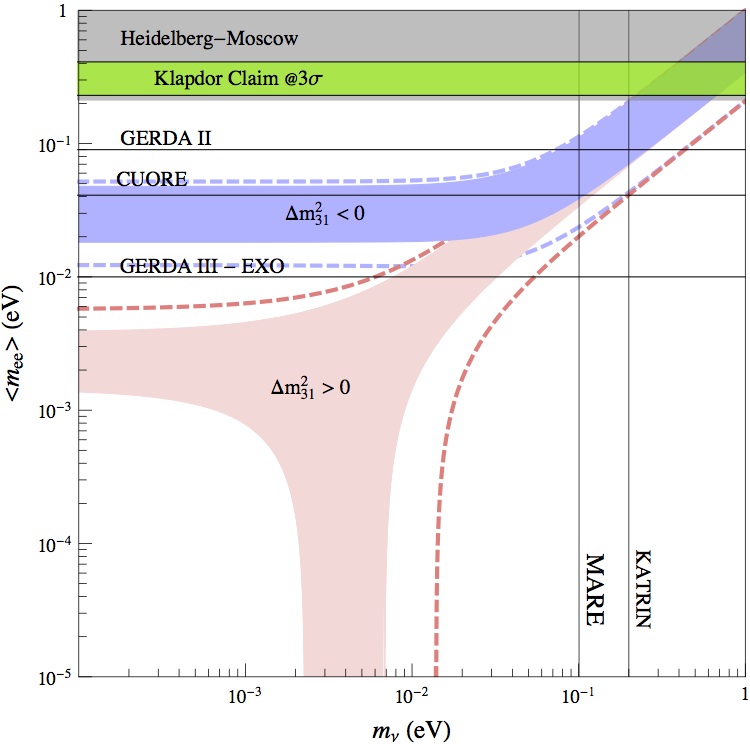}}
\subfigure[Correlations.]
{\includegraphics[width=7cm,height=7cm]{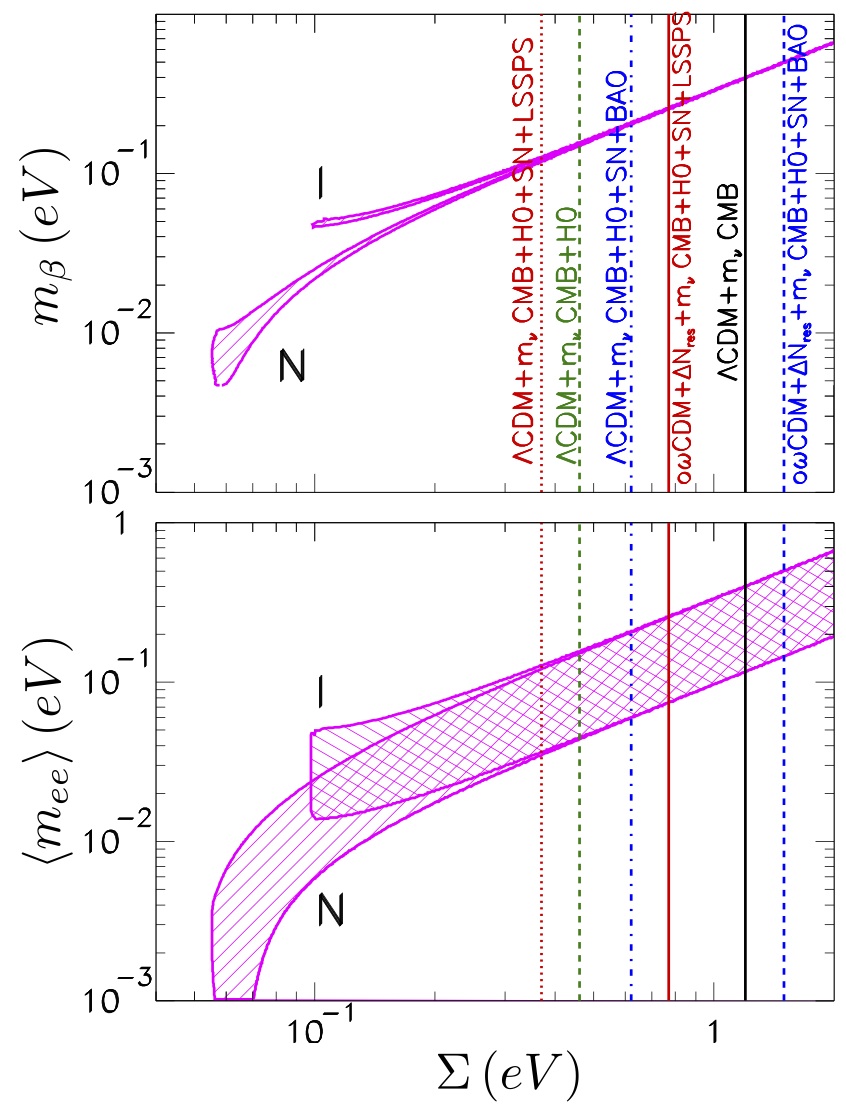}}
\caption{(a) The $0\nu2\beta$ effective mass as a function of the lightest neutrino mass for both the mass orderings: in Red (Blue) the NO (IO). The coloured areas (dashed lines) correspond to the neutrino oscillation parameters at $1\sigma$ ($3\sigma$) from Ref.~\cite{Fogli:2012ua}. (b) Correlation between $\Sigma$ and $m_\beta$ and $\langle m_{ee}\rangle$. See details in the text.}
\label{fig:0nu2betaAndCorrelations}
\end{figure}

Taking the two Majorana phases in the range $[0,\pi]$, one can generate the plot in Fig.~\ref{fig:0nu2betaAndCorrelations}(a), where the $0\nu2\beta$ effective mass is drown as a function of the lightest neutrino mass for both the mass orderings. The horizontal lines corresponds to the past and future sensitivities of the $0\nu2\beta$ decay experiments: the Hidelberg-Moscow experiment \cite{KlapdorKleingrothaus:2000sn} put an upper bound at $90\%$ of C.L. of $\mean{m_{ee}}<0.21-0.53\,\text{eV}$, while a part of the collaboration claimed the observation of the $0\nu2\beta$ decay \cite{KlapdorKleingrothaus:2006ff} corresponding to a value $\mean{m_{ee}}=0.32\pm0.03\,\text{eV}$; the expected future sensitivities of GERDA II \cite{Abt:2004yk}(GERDA III, EXO\cite{Danilov:2000pp}) are of $\mean{m_{ee}}\sim90-150\,(10)\,\text{eV}$, while that one of CUORE \cite{Arnaboldi:2003tu} is of $\mean{m_{ee}}\sim41-96\,(10)\,\text{eV}$. The vertical ones to the future sensitivity of KATRIN \cite{Host:2007wh}, $0.2\;\text{eV}$, and MARE \cite{Monfardini:2005dk}, 0.1\;\text{eV}.

In Fig.~\ref{fig:0nu2betaAndCorrelations}(b), correlations among the sum of the neutrino masses, the single $\beta$ parameter and the $0\nu2\beta$ effective mass are shown. The vertical lines show the recent bounds on $\Sigma$ \cite{GonzalezGarcia:2010un} considering two contexts and different combination of the cosmological data: the Blue corresponds to the standard cosmological model $\Lambda$CDM with massive neutrinos and the Red the generalization with non-vanishing curvature and with $\omega\neq-1$ in the Dark Matter equation of state. CMB stands for the Cosmic Microwave Background, HO for the Hubble Constant, SN for the high-redshift Type-I SuperNovae, BAO for the Baryon Acoustic Oscillation, LSSPS for Large Scale Structure matter Power Spectrum.

Considering a specific model and the corresponding sum rule, it is possible to restrict the parameter space of the three observables: indeed the two Majorana phases enter the sum rule and prevent the determination of a single point in their parameter space. Studies on such bounds in specific models have been presented in Refs.~\cite{Barry:2010yk,Dorame:2011eb,Dorame:2012zv}, where it has been shown how the implementation of the sum rules could represent a strong improvement to test flavour models for neutrino masses and mixings. 

However, as also pointed out in Ref.~\cite{Barry:2010yk}, these correlations among the neutrino masses are perturbed due to the presence of the higher order operator contributions. As a result, it is not possible to identify a precise point in the parameter space $(\Sigma,m_\beta)$ or a precise lower bound on $\mean{m_{ee}}$, the uncertainty being proportional to the amount of the corrections. For the FD mixing patterns discussed in the previous sections the required correction to the reactor angle to suitably reproduce the corresponding experimental value must be of the order of $\cO(0.1)$. As a result, the relative corrections to the neutrino masses are expected to be of the same order of magnitude, that introduce a sensible uncertainty to the sum rule and the corresponding constraints on $\Sigma$, $m_\beta$ and $\mean{m_{ee}}$.

\subsection{Lepton Flavour Violation}

In this section we discuss the implications for lepton flavour violating (LFV) processes of the classes of models described in the previous sections. In particular, we will focus on radiative lepton decays, that turn out to give very strong constraints.

For all the effective models described here, we will consider a working framework in which SUSY is broken at a high-scale, higher than or comparable to the flavour breaking one $\Lambda_f$, in a hidden sector. Subsequently, this breaking is transmitted at the lower scale $\Lambda$ of few TeV by a mechanism such as gravity mediation. In this context, the soft SUSY breaking terms undergo the flavour symmetry and specific non-universal BCs arise at the scale $\Lambda_f$. It is the non-universality of the BCs that differentiates the contributions to the radiative lepton transitions of the flavour models under discussion from the well-known MSUGRA framework.

\subsubsection{The SUSY Lagrangian and the Soft SUSY Breaking Terms}

The Lagrangian of a generic SUSY model is given by
\beq
\LL=\int d^2\Theta d^2\overline{\Theta}\, \cK(\ov{z}, e^{2 V} z)+\left[\int d^2 \Theta\, w(z)+\hc\right]+
     \dfrac{1}{4}\left[\int d^2\Theta\, f(z) \,\cW\,\cW+\hc\right]\,,
\label{leel}
\eeq
where $\cK(\ov{z},z)$ is the K\"ahler potential, $w(z)$ is the superpotential and $f(z)$ is the gauge kinetic 
function. $V$ is the Lie-algebra valued vector supermultiplet, describing the gauge fields and their superpartners 
while $\cW$ is the chiral superfield describing, together with the function $f(z)$, the kinetic terms of gauge 
bosons and their superpartners. 

When the flavour symmetry is broken, the kinetic terms could receive non-canonical contributions. In totally general notation, we can write
\beq
\LL_{kin}=i\,K_{ij}\,\bar \ell_i\,\bar\sigma^{\mu}\,D_\mu\,\ell_j+i\,K^c_{ij}\,\bar \ell^c_i\bar\sigma^{\mu}\,D_\mu\,\ell_j^c+K_{ij}\,\ov{D^\mu\,\tilde \ell_i}\,D_\mu\,\tilde \ell_j+K^c_{ij}\,\ov{D^\mu\tilde \ell^c_i}\,D_\mu\,\tilde \ell^c_j\,,
\label{Lkinetic}
\eeq
where $D_\mu$ is the covariant derivative and $K^{(c)}$ are hermitian $3\times3$ matrices in the flavour space.

The superpotential can be written distinguishing several parts:
\beq
w=w_\ell+w_\nu+w_d+w_h\,,
\eeq
where, once the flavour symmetry is broken, $w_\ell$ ($w_\nu$) is responsible for the charged lepton (neutrino) masses, $w_d$ contains the self-interactions of the flavons and possibly their interactions with other superfields, apart from those already described in $w_\ell$ and $w_\nu$ and with the Higges. Finally, the last term $w_h$ is associated to the $\mu$ term
\beq
w_h=x_\mu\,H_u\,H_d\,.
\eeq

The soft SUSY breaking terms are generated from this SUSY Lagrangian analytically continuing all the coupling constants (such as the couplings in the superpotential and in the K\"ahler potential) into superspace, i.e. by promoting all the coupling constants to superfields with constant $\Theta^2$ and $\Theta^2\bar\Theta^2$ components \cite{Giudice:1997ni,Luty:2005sn}. In particular, the Lagrangian describing the slepton masses is
\beq
-\LL_m\supset\quad
\left(\ov{\tilde{e}}\quad\tilde{e}^c\right)
\cM_e^2
\left(
  \begin{array}{c}
    \tilde{e} \\
    \ov{\tilde{e}}^{c} \\
  \end{array}
\right)+
\ov{\tilde{\nu}}\, m^2_{\nu LL}\,\tilde{\nu}\,,
\eeq
where
\beq
\cM_e^2=\left(
  \begin{array}{cc}
    m^2_{eLL} & m_{eLR}^2 \\[1mm]
    m_{eRL}^2 & m^2_{eRR} \\
  \end{array}
\right)\,,
\eeq
with $m^2_{(e,\nu)LL}$ and $m^2_{eRR}$ being $3\times 3$ hermitian matrices, while $m^2_{eLR} = 
\left(m^2_{eRL}\right)^\dag$. Each of these blocks receives contributions from different part of 
the SUSY Lagrangian:
\beq
\begin{aligned}
m^2_{(e,\nu)LL}&=(m^2_{(e,\nu) LL})_K+(m^2_{(e,\nu) LL})_F+(m^2_{(e,\nu) LL})_D\,,\\
m^2_{eRR}&=(m^2_{eRR})_K+(m^2_{eRR})_F+(m^2_{eRR})_D\,,\\
m^2_{eRL}&=(m^2_{eRL})_1+(m^2_{eRL})_2 \,.
\end{aligned}
\eeq
The indexes specify the origin of these terms: $K$ stands for the K\"abler potential; $F$ and $D$ for the SUSY $F-$ and $D-$term contributions; $(m^2_{eRL})_1$ refers to the contributions originated by the analytically continuation into superspace of the coupling constants of the superpotential, while $(m^2_{eRL})_2$ to the usual term proportional to the $\mu$ parameter. Notice that in the sneutrino sector only the LL block is present and that any contribution to the sneutrino masses associated to $w_\nu$ can be safetely neglected. 

In order to compute the contributions to the LFV transitions, it is first necessary to move to the {\it physical basis}, where the kinetic terms are canonical and the charged lepton mass matrix is diagonal. Performing the same transformations in all the components of the interested chiral supermultiplets, to avoid flavour-violating gaugino-lepton-slepton vertices, it is possible to recover the complete set of non-universal BCs at the scale $\Lambda_f$. Since the radiative lepton decays occur at lower-energies than $\Lambda_f$, it is necessary to evolve the soft SUSY breaking terms down to these energies, considering the non-universal BCs.

We will specify the flavour-dependent contributions for the slepton mass matrices in the physical basis in the following, when entering more in the details of the different classes of models, while here we report the supersymmetric $F-$ and $D-$ term contributions and $(m^2_{eRL})_2$:
\begin{align}
&(m_{eLL}^2)_F=m_\ell^T m_\ell  && (m^2_{eLL})_D=\left(-\frac{1}{2}+\sin^2\theta_W \right) \cos 2\beta ~m_Z^2 \times \unity\\
&(m_{\nu LL}^2)_F=0 					&&(m^2_{\nu LL})_D=\left(+\frac{1}{2} \right) \cos 2\beta~m_Z^2 \times \unity\\
&(m_{eRR}^2)_F=m_\ell m_\ell^T &&(m^2_{eRR})_D=-\sin^2\theta_W \cos 2\beta ~m_Z^2 \times \unity\\[2mm]
&(m^2_{eRL})_2=-\mu \tan\beta~ m_\ell\,.
\end{align}

\subsubsection{The LFV Transitions} 

Having defined the procedure to get the low-energy expressions for the soft SUSY breaking terms, we can now proceed with the radiative lepton decays, $\mu\to e \gamma$, $\tau\to\mu\gamma$ and $\tau\to e \gamma$. In this section we provide the basic formulae for the computation of the normalized branching ratio $R_{ij}$ for these LFV transitions in the so-called mass insertion (MI) approximation \cite{Paradisi:2005fk} (for the complete formulae see for example Refs.~\cite{Hisano:1995cp,Fukuyama:2005bh,Arganda:2005ji}). In this approximation, the observables are expressed in terms of the off-diagonal elements of the slepton mass matrices $m_{SUSY}$, normalized to their average mass, and the normalized branching ratio $R_{ij}$ can be expressed as:
\beq
R_{ij}= \frac{48\pi^3 \alpha}{G_F^2 m_{SUSY}^4}
\left(\vert A_L^{ij} \vert^2+\vert A_R^{ij} \vert^2 \right)\,.
\label{rij}
\eeq
At the LO, the amplitudes $A_L^{ij}$ and $A_R^{ij}$ are given by:
\beq
\begin{aligned}
A_L^{ij}&=&a_{LL} (\delta_{ij})_{LL} + a_{RL} \frac{m_{SUSY}}{m_i} (\delta_{ij})_{RL}\\
A_R^{ij}&=&a_{RR} (\delta_{ij})_{RR} + a_{LR} \frac{m_{SUSY}}{m_i} (\delta_{ij})_{LR}
\end{aligned}
\label{ALAR}
\eeq
with $a_{CC'}$ $(C,C'=L,R)$ dimensionless functions of the SUSY parameters. We list in Tab.~\ref{table_aCC} the expressions and the numerical values of the functions $a_{CC'}$, in the limit $\mu=M_{1,2}=m_{SUSY}$. The parameters $\delta_{CC'}$ are the mass insertions of the $m^2_{eCC'}$:
\beq
\left(\delta_{ij}\right)_{CC'}=\dfrac{\left(m^2_{eCC'}\right)_{ij}}{m^2_{SUSY}}\,.
\eeq

\begin{vchtable}[h!]
\vchcaption{Coefficients $a_{CC'}$ characterizing the transition amplitudes for $\mu\to e \gamma$, $\tau\to e \gamma$ and $\tau\to \mu\gamma$, in the MI approximation, taking the limit the $\mu=M_{1,2}=m_{SUSY}$ and universal BCs. Numerical values are given in units of $g^2/(192 \pi^2)$ and for $\tan\beta=(2,20)$.}
\label{table_aCC} 
\begin{tabular}{@{}ccc@{}}
\hline
&&\\[-3mm]
$a_{LL}$ & $\dd\frac{1}{240}\frac{g^2}{16 \pi^2}\left[1-15\tan^2\theta_W+4 \Big(4+5\tan^2\theta_W\Big)\tan\beta\right]$ & $+(2,\,22)$ \\[3mm]
\hline
&&\\[-3mm]
$a_{RL}=a_{LR}$ & $\dd\frac{1}{12}\frac{g^2}{16 \pi^2}\tan^2\theta_W$ & $0.30$ \\[3mm]
\hline
&&\\[-3mm]
$a_{RR}$ & $\dd\frac{1}{60}\frac{g^2}{16 \pi^2}\tan^2\theta_W\left[-6-\tan\beta\right]$ & $-(0.5,\,1.6)$ \\[3mm]
\hline
\end{tabular}
\end{vchtable}

It is interesting now to discuss the main sources of flavour violation that can affect these observables. We first consider the case without RH neutrinos, for which the RGE running has a negligible effect to the low-energy expressions of the soft SUSY breaking terms, as it is discussed in the App.~\ref{AppRGE}. In this case, it is easy to identify the source of flavour violation and to estimate its contributions. 
\begin{itemize}
\item[$(\delta_{ij})_{LL,RR}$:] Contributions to the off-diagonal entries of $(\delta_{ij})_{LL}$ arise in correspondence of non-canonical kinetic terms due to multiple flavon insertions. Indeed, in the K\"ahler potential, it is always possible to introduce a term as $\Phi^\dag \Phi$, with $\Phi$ a generic flavon, and typically this term produces off-diagonal entries of the order of $VEV^2/\Lambda_f^2$. The off-diagonal entries of $(\delta_{ij})_{LL}$ are generically proportional to the same quantities. Similar comments apply to the $(\delta_{ij})_{RR}$ term, but its impact on the ratios $R_{ij}$ is subdominant with respect to $(\delta_{ij})_{LL}$, as follows from Tab.~\ref{table_aCC}. In some cases, when the flavor $\Phi$ transforms only under $S_4$ and in particular is a singlet under the additional Abelian symmetries that define $G_f$, it is possible to generate $d=3$ terms in the K\"ahler potential. As a result the off-diagonal entries could be proportional to $VEV/\Lambda_f$. This situation is indeed realized in the models in Refs.~\cite{Altarelli:2009gn,Toorop:2010yh,Meloni:2011fx}, dealing with the BM pattern.

\item[$(\delta_{ij})_{RL,LR}$:] The off-diagonal entries of $(\delta_{ij})_{RL}$ (and $(\delta_{ij})_{LR}$) in the flavour basis have the same origin of the charged lepton Yukawa entries. In the mass basis, it is unlucky a change of the order of magnitudes of the off-diagonal entries: indeed, this can happen only in the case of precise alignments in the flavour space of the contributions to the Yukawa matrix and the slepton mass matrix. The flavour structure of the charged lepton Yukawa, and of the RL block of the slepton masses, strongly depends on the class of models. Notice that the contributions associated to $(\delta_{ij})_{RL}$ are dominant over those associated to $(\delta_{ij})_{LR}$, due to a suppressing factor $m_j/m_i$.
\end{itemize}
Whether the $(\delta_{ij})_{LL}$ contributions or the $(\delta_{ij})_{RL}$ ones are the dominant one in the ratios $R_{ij}$ depends on the magnitude of $VEV/\Lambda_f$ and of the off-diagonal entries of the charged lepton Yukawas: both of these aspects are model dependent and we will enter in the details of the different class of models in the next section.

When RH neutrinos are present in the spectrum then the direct connection between the contributions in the Lagrangian and the flavour violating effects is partially lost. Indeed, from closer look to the RGEs (eqs.~(\ref{RGEmLLRH})-(\ref{RGEYeRH})), it follows that large $(\delta_{ij})_{LL}$ are generated even in the case of universal BCs, due to the $\cO(1)$ entries of the Dirac neutrino Yukawa. Depending on the specific lepton mixing and on the flavour BCs of a model, 1) it is possible to neglect the specific flavour BCs and $(\delta_{ij})_{LL}$ is determined only by the RGE evolution with universal BCs, or 2) $(\delta_{ij})_{LL}$ results from the RGE evolution considering the flavour BCs. We will discuss further these possibilities in the next section, when specifying the class of models.

\boldmath    
\subsubsection{LFV in TB Models} 
\unboldmath

For models with the TB mixing, as discussed in Sec.~\ref{sec:NuPatterns}, focussing only to the scenario with the highest success rate and only on the NH case, when the parameter $\theta$ takes the value of $0.086$ then the success rate to get the mixing angles inside their present $3\sigma$ ranges is maximized. For such value the contributions to flavour violating radiative lepton transitions are indeed relevant. 

In order to illustrate the general predictions of the TB models, we assume the following mass matrices for the sleptons, in the physical basis:
\begin{align}
({m}_{(e,\nu)LL}^2)_K &=
\left( \begin{array}{ccc}
                n_0+ n_1\,\theta^2  		& n_4\,\theta^2 							&  n_5\,\theta^2 \\
                n_4\,\theta^2 					& n_0+ n_2\,\theta^2					&  n_6\,\theta^2 \\
                n_5\,\theta^2  					& n_6\,\theta^2 							& n_0+ n_3\,\theta^2 \\
\end{array}\right) \, m_0^2\\
({m}_{eRR}^2)_K &= \left( \begin{array}{ccc}
                n^c_1  																				
                &  n_4^c\,\dfrac{m_e}{m_\mu}\theta					
                &  n_5^c\,\dfrac{m_e}{m_\tau}\theta	\\[3mm]
                n_4^c\,\dfrac{m_e}{m_\mu}\theta			
                &  n^c_2 																						
                &  n_6^c\,\dfrac{m_\mu}{m_\tau} \,\theta\\[3mm]
                n_5^c\,\dfrac{m_e}{m_\tau}\theta	 			
                &  n_6^c\,\dfrac{m_\mu}{m_\tau} \,\theta		
                &  n^c_3 \\
	\end{array}\right) \, m_0^2\\
({m}_{eRL}^2)_1 &=
\left( \begin{array}{ccc}
                	z_e\,m_e                							
                	& z'_e\,m_e\,\theta																				
                	& z^{\prime\prime}_e\,m_e\,\theta \\[3mm]
                	z'_\mu\,m_\mu\,\theta				
                	& z_\mu\,m_\mu 
                	& z^{\prime\prime}_\mu\,m_\mu\,\theta\\[3mm]
                	z'_\tau\,m_\tau\,\theta               
                	& z^{\prime\prime}_\tau\,m_\tau\,\theta
		       	& z_\tau\,m_\tau \\
	\end{array}\right) \, A_0
\end{align}
where all the parameters should be consider as random complex numbers with absolute value $\cO(1)$, apart from $n_0$ and $n_i^c$ that are positive to favour positive definite square-masses and to avoid electric-charge breaking minima and further sources of electroweak symmetry breaking. 

The $LL$ contributions provide the dominant effects on the branching ratio of the lepton radiative transitions for larger values of the $\tan\beta$, while for small values also the $RL$ contributions turn out to be relevant. Furthermore, in absence of RH neutrinos, the RGE evolution do not give relevant effects to the low-energy expressions of the slepton mass matrix. As a result, entering these expression for the slepton mass matrices in eq.~(\ref{ALAR}), we find
\beq
R_{ij}\simeq\frac{48\pi^3 \alpha}{G_F^2 m_{SUSY}^4}\left(\left| a_{LL}\, \cO(\theta^2)\right|^2+\left| a_{RL}\, \cO(\theta)\right|^2 \right)\, 
\eeq
for all the three transitions.

\begin{figure}[h!]
 \centering
\subfigure[$m_0=200$ GeV, $\tan\beta=2$.]
   {\includegraphics[width=7cm]{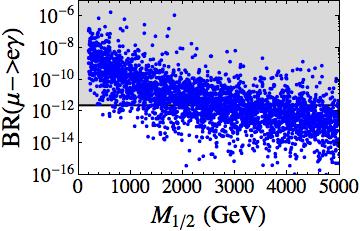}}
\subfigure[$m_0=5000$ GeV, $\tan\beta=2$.]
   {\includegraphics[width=7cm]{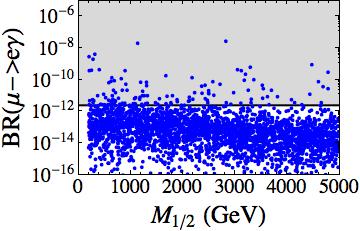}}
\subfigure[$m_0=200$ GeV, $\tan\beta=20$.]
   {\includegraphics[width=7cm]{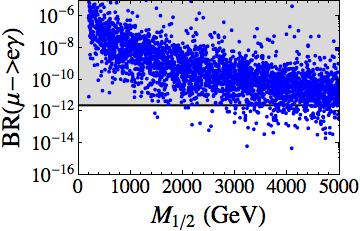}}
\subfigure[$m_0=5000$ GeV, $\tan\beta=20$.]
   {\includegraphics[width=7cm]{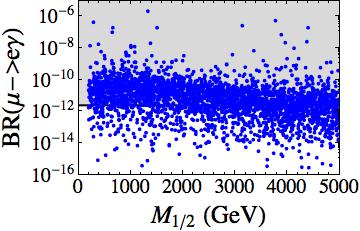}}
\caption{ {\bf TB Models.} Scatter plots of $BR(\mu\to e \gamma)$ as a function of $m_{1/2}$, for $\tan\beta=(2,20)$, $\theta=0.103$ and different values for $m_0$. The horizontal lines show the current MEG bound $BR(\mu\to e \gamma)\lesssim2.4\times 10^{-12}$.}
\label{fig:MEGTBWeinberg}
\end{figure}

When considering the present bound on $BR(\mu\to e\gamma)$ from the MEG collaboration, we can constrain the supersymmetric parameter space of the model. The results are shown in Fig.~\ref{fig:MEGTBWeinberg}, that represents the dependence of $BR(\mu\to e\gamma)$ on $M_{1/2}$ for fixed values of $m_0$. The value for $\theta$ in these plots is $0.103$, that maximizes the success rate when the corrections to the TB pattern arise from the charged lepton sector, which has the best statistical chances. Only the NH case is shown in the plots. The results for the IH case as well as those when $\theta=0.104$ are very similar.

When considering models with RH neutrinos, the analysis is slightly different and it is necessary to account for the RGE contributions. Considering the flavour BCs in Eqs.~(\ref{meLLS4BM})-(\ref{meRLS4BM}), we find that the impact on the RG evolution into all the three branching ratios is relevant. In particular, at low-energy, the off-diagonal entries of $\left(m^2_{eLL} \right)_K$ are multiplied by a factor 3(2), for the NH (IH) case, with respect to the corresponding entries at the high-energy. However, this does not change the prediction that 
\beq
R_{\mu\,e}\approx R_{\tau\,e}\approx R_{\tau\,\mu}\,,
\eeq
and also the the plots in Fig.~\ref{fig:MEGTBWeinberg} can safely apply in the case with RH neutrinos.

\boldmath    
\subsubsection{LFV in BM Models} 
\unboldmath

In this class of models, the lepton mixing matrix at LO coincides with the BM pattern and larger corrections are necessary to bring the solar angle in agreement with the data. In the models presented in Refs.~\cite{Altarelli:2009gn,Toorop:2010yh,Patel:2010hr,Meloni:2011fx}, such corrections arise from the charged lepton sector and are of the order of $\theta\approx\cO(0.1)$, as we have previously discussed. No model has been presented so far in which the corrections arise from the neutrino sector and therefore we disregard this possibility in the following analysis.

In this context and considering for the next analysis the specific case of Ref.~\cite{Altarelli:2009gn}, the particular charge assignment of the flavons and their vacuum alignment determine an hierarchy between the off-diagonal entries of the slepton mass matrices. Moreover, a triplet of $S_4$ not transforming under the additional Abelian symmetries of $G_f$ is present in the model and as a result it originates non-renormalisable terms in the K\"ahler potential suppressed by only one power of the cut-off scale $\Lambda_f$. More in details, in the physical basis we get 
\begin{align}
({m}_{(e,\nu)LL}^2)_K &=
\left( \begin{array}{ccc}
                n_0+ n_1\,\theta  		&  n_2\,\theta 						&  n_2\,\theta \\
                n_2\,\theta 				& n_0+ n_1\,\theta				&  n_3\,\theta^2 \\
                n_2\,\theta  				& n_3\,\theta^2 						& n_0+ n_1\,\theta \\
\end{array}\right) \, m_0^2
\label{meLLS4BM}\\
({m}_{eRR}^2)_K &= \left( \begin{array}{ccc}
                n^c_1  												&  n_4^c\,\dfrac{m_e}{m_\mu}\,\theta				&  n_5^c\,\dfrac{m_e}{m_\tau}\,\theta\\[3mm]
                n_4^c\,\dfrac{m_e}{m_\mu}\,\theta 		&  n^c_2 														&  n_6^c\,\dfrac{m_\mu}{m_\tau} \,\theta^2\\[3mm]
                n_5^c\,\dfrac{m_e}{m_\tau}\,\theta  	&  n_6^c\,\dfrac{m_\mu}{m_\tau} \,\theta^2	&  n^c_3 \\
	\end{array}\right) \, m_0^2
\end{align}
\beq
({m}_{eRL}^2)_1 =
\left( \begin{array}{ccc}
                z_e\,m_e                					& z'_e\,m_e\,\theta  										& z^{\prime\prime}_e\,m_e\,\theta \\[3mm]
                z'_\mu\,m_\mu\,\theta				& z_\mu\,m_\mu 			            						& z^{\prime\prime}_\mu\,m_\mu\,\theta^2\\[3mm]
                z'_\tau\,m_\tau\,\theta               	& z^{\prime\prime}_\tau\,m_\tau\,\theta^2      & z_\tau\,m_\tau \\
	\end{array}\right) \, A_0
\label{meRLS4BM}
\eeq
where all the parameters are treat as in the previous case with TB models.

The $LL$ contributions provide the dominant effects on the branching ratio of the lepton radiative transitions. Also for this case, in absence of RH neutrinos, the RGE evolution do not give relevant effects to the low-energy expressions of the slepton mass matrix. As a result, entering these expression for the slepton mass matrices in eq.~(\ref{ALAR}), we find
\beq
\begin{aligned}
&R_{\mu e}=R_{\tau e}\simeq\frac{48\pi^3 \alpha}{G_F^2 m_{SUSY}^4}\left(\left| a_{LL} \right|^2+\left| a_{RL} \right|^2 \right)\, \left|\cO(\theta)\right|^2\,,\\
&R_{\tau\mu}\simeq\frac{48\pi^3 \alpha}{G_F^2 m_{SUSY}^4}\left(\left| a_{LL} \right|^2+\left| a_{RL} \right|^2 \right)\, \left|\cO(\theta^2)\right|^2\,,
\end{aligned}
\eeq
and therefore
\beq
R_{\mu e}\simeq R_{\tau e}\gg R_{\tau\mu}\,.
\label{BRsBM1}
\eeq

\begin{figure}[h!]
 \centering
\subfigure[$m_0=200$ GeV, $\tan\beta=2$.]
   {\includegraphics[width=7cm]{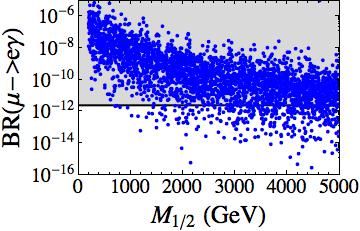}}
\subfigure[$m_0=5000$ GeV, $\tan\beta=2$.]
   {\includegraphics[width=7cm]{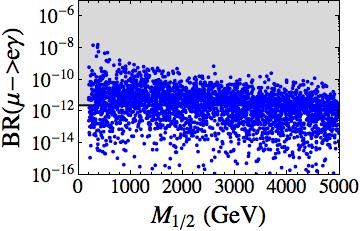}}
\subfigure[$m_0=200$ GeV, $\tan\beta=20$.]
   {\includegraphics[width=7cm]{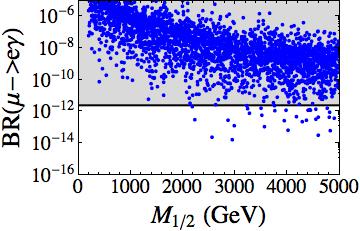}}
\subfigure[$m_0=5000$ GeV, $\tan\beta=20$.]
   {\includegraphics[width=7cm]{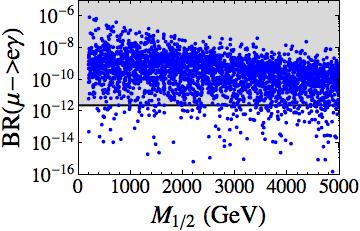}}
\caption{{\bf BM Models.} Scatter plots of $BR(\mu\to e \gamma)$ as a function of $m_{1/2}$, for $\tan\beta=(2,20)$, $\theta=0.156$ and different values for $m_0$. The horizontal lines show the current MEG bound $BR(\mu\to e \gamma)\lesssim2.4\times 10^{-12}$.}
\label{fig:MEGBMWeinberg}
\end{figure}

As for the previous case with the TB mixing, the most stringent constraint come from the $BR(\mu\to e\gamma)$ and the results are shown in Fig.~\ref{fig:MEGBMWeinberg}. The plots represent the dependence of $BR(\mu\to e\gamma)$ on $M_{1/2}$ for fixed values of $m_0$. The value for $\theta$ in these plots is $0.156$, that maximizes the success rate when the corrections to the BM pattern arise from the neutrino sector, which has the best statistical chances. Only the NH case is shown in the plots. The results for the IH case as well as those considering $\theta=0.157$ are very similar.

When considering models in which RH neutrinos enrich the spectrum, the analysis can be different. Considering the flavour BCs in Eqs.~(\ref{meLLS4BM})-(\ref{meRLS4BM}), we find that the impact on the RG evolution into $R_{\mu\,e}$ and $R_{\tau\,e}$ is negligible and that only into $R_{\tau\,\mu}$ is relevant. In particular, at low-energy, the off-diagonal entries of $\left(m^2_{eLL} \right)_K$ are all of order of $\cO(\theta)$. As a result the previous prediction for the branching ratios in eq.~(\ref{BRsBM1}) is modified and considering the presence of RH neutrinos it is given by:
\beq
R_{\mu\,e}\approx R_{\tau\,e}\approx R_{\tau\,\mu}\,.
\eeq
Considering more in detail the $\mu\to e \gamma$ transition, the plots in Fig.~\ref{fig:MEGBMWeinberg} also apply in the case with RH neutrinos.

\section{Conclusions}
\label{sec:Concl}

In the last decade, numerous flavour models based on discrete symmetries have shown their ability in accommodating the lepton mixings. In particular, special mixing patterns, that predict the mixing angles independently of the neutrino spectrum, are achieved at the LO: these patterns are the TB, the BM, the GR and some TM schemes, that we refer to as FD patterns.

Not with all the discrete symmetries commonly used it is possible to describe all these special mixing patterns. However, the group $S_4$ has been implemented in different models in order to describe both the TB, the BM and the TM schemes. In this paper, we have reviewed the main features of the $S_4$ discrete group, summarized on the existing models based on this group and analyzed general phenomenological signatures of these models.

A relevant question that we accounted for is about the naturalness of the FD patterns, after that the reactor angle has been proven to be non-vanishing. In particular, we analyzed the success rates of TB and the BM patterns with general corrections arising either from the charged lepton sector or by the neutrino sector. The results favour the TB scheme with general corrections from the charged lepton sector. For this case, the success rate reaches about the $15\%$. Better results could be achieved considering specific type of corrections, that for example affect only two of the mixing angles and not all of them. However, in this case, the success of the model is due to a dynamical trick, consisting in reducing as much as possible the number of NLO independent corrections, that, moreover, must act in specific directions in the flavour space.

Apart the naturalness requirement, a complete model should undergo to a series of criteria: account for a mechanism for the flavour symmetry breaking; describe the quark sector in a unified approach, i.e. using the same flavour symmetry both for quarks and leptons; produce predictions in the lepton and quark sectors. More criteria are satisfied more a model should be considered complete even if it is true that nature could have been chosen an unexpected way.

In order to test these models and to distinguish one model from the others, it is necessary to investigate on their phenomenological signatures. We pointed out that often sum rules are present in discrete flavour models and as a result quite clear correlations arise among the sum of the neutrino masses, the kinematic electron mass in the beta decay and the neutrino-less double beta decay effective mass. Alternatively, lepton flavour violating transitions, such as $\ell_i\to\ell_j\gamma$, provide strong constraints, once assuming the presence of flavour-sensitive new physics at the TeV scale. In this paper, we concentrated in the supersymmetric case, distinguishing the case in which the neutrino masses are explained through the Weinberg operator or through the type I See-Saw mechanism. Even if the contributions from the RGE are different in the two cases, the constraints arising form the $\mu\to e\gamma$ decay are similar: these models are not compatible with very light supersymmetry.

Even though the huge effort of these years in constructing flavour models to describe masses and mixings for the neutrinos, and more in general for all the fermions, it is discouraging that no illuminating strategy arise form this scenario. On the other hand, this is partially related to the large uncertainties still present in the flavour sector. The hope is that with a better determination of the lepton mixing angles and with the knowledge of the CP phases, the neutrino mass scale, the type of the neutrino nature and spectrum, it will be finally possible to shed light on the origin of the fermion masses and mixings.

\begin{acknowledgement}
LM recognises that this work has been partly supported by the Te\-ch\-ni\-sche Universit\"at M\"unchen -- Institute for Advanced Study, funded by the German Excellence Initiative.
\end{acknowledgement}


\appendix

\section{Renormalisation Group Effects} 
\label{AppRGE}

The renormalization group equations in the MSSM context for the soft mass terms $(m^2_{(e,\nu)LL} )_{K}$,  $(m^2_{eRR} )_{K}$ and $A_e \equiv \sqrt{2}(m^2_{eRL})_{1}/(v \cos \beta)$, denoting $ t' \equiv \log (\Lambda_L/m_{SUSY})$, are \cite{Martin:1993zk,Hisano:1995cp}:
\begin{align}
\begin{split}
16\pi^2 \mu\frac{d}{d \mu} \left( m^2_{eLL} \right)_{K_{ij}} =&
-\left( \frac{6}{5} g_1^2 \left| M_1 \right|^2 + 6 g_2^2 \left| M_2 \right|^2 \right) \delta_{ij} -\frac{3}{5} g_1^2~S~\delta_{ij}\\
&+  \left (( m^2_{eLL} )_{K} Y_e^{\dagger} Y_e + Y_e^{\dagger} Y_e ( m^2_{eLL} )_{K} \right)_{ij} \\
&+ 2 \left( Y_e^{\dagger} (m^2_{eRR} )_{K} Y_e +{m}^2_{H_d} Y_e^{\dagger} Y_e +A_e^{\dagger} A_e \right)_{ij}\,,
\end{split}
\label{RGEmLL}\\
\begin{split}
16\pi^2 \mu\frac{d}{d \mu} \left( m^2_{eRR} \right)_{K_{ij}} =&
- \frac{24}{5} g_1^2 \left| M_1 \right|^2 \delta_{ij} + \frac{6}{5} g_1^2~S~\delta_{ij} \\
&+ 2 \left ( ( m^2_{eRR} )_{K}  Y_e Y_e^{\dagger} + Y_e Y_e^{\dagger} ( m^2_{eRR} )_{K} \right)_{ij}\\
&+ 4 \left( Y_e ( m^2_{eLL} )_{K} Y_e^{\dagger} + {m}^2_{H_d}Y_e Y_e^{\dagger} +  A_e A_e^{\dagger} \right)_{ij}\, 
\end{split}
\label{RGEmRR}\\
\begin{split}
16\pi^2 \mu\frac{d}{d \mu}  A_{e_{ij}} =&
 \left( -\frac{9}{5} g_1^2 -3 g_2^2+ 3 {\rm Tr} ( Y_d^{\dagger} Y_d )+   {\rm Tr} ( Y_e^{\dagger} Y_e ) \right)A_{e_{ij}}\\
&+ 2 \left(\frac{9}{5} g_1^2 M_1 + 3 g_2^2 M_2+ 3 {\rm Tr} ( Y_d^{\dagger} A_d)+   {\rm Tr} ( Y_e^{\dagger} A_e) \right) Y_{e_{ij}} \\
&+ 4 \left( Y_e Y_e^{\dagger} A_e \right)_{ij}+ 5 \left(A_e Y_e^{\dagger} Y_e \right)_{ij}\,,
\end{split}
\label{RGEAe}\\
16\pi^2 \mu\frac{d}{d \mu} Y_{e_{ij}} = &
\left ( -\frac{9}{5} g_1^2 - 3 g_2^2 + 3 \,{\rm Tr} ( Y_d Y_d^{\dagger}) +   {\rm Tr} ( Y_e Y_e^{\dagger})\right ) Y_{e_{ij}}+3 \, \left( Y_e Y_e^{\dagger} Y_e \right)_{ij} \,,
\label{RGEYe}
\end{align}
where $g_{1,2}$ are the gauge couplings \footnote{In the GUT normalization, such that $g_2=g$ and $g_1=\sqrt{5/3} g'$.} of SU(2)$_L\times U(1)_Y$, $M_{1,2}$ the corresponding gaugino mass terms, $Y_{e,d}\equiv m_{\ell,d}\,/\,v_d$ are the Yukawa matrices for charged leptons and down quarks, $A_d=(m^2_{dRL})_{1}\,/\,v_d$ and:
\begin{equation}
S = {\rm Tr} \left(m^2_{qLL} + m^2_{dRR}- 2 m^2_{uRR}- (m^2_{eLL})_K + (m^2_{eRR})_K \right) - {m}^2_{H_d}+ {m}^2_{H_u}\,.
\end{equation}
The matrix $(m^2_{\nu LL})_K$ coincides with $(m^2_{e LL})_K$ and has the same evolution.
For squarks we have introduced soft mass terms analogous to those previously discussed for sleptons. 
To estimate the corrections to the slepton masses induced by the renormalization group evolution we adopt the leading logarithmic approximation and
substitute each of the running quantities with their initial conditions at the scale $\Lambda_L\approx\Lambda_f$ in eqs.~(\ref{RGEmLL})--(\ref{RGEYe}). 

In this approximation and considering that in general the off-diagonal entries of the soft mass matrices are smaller than the corresponding elements into the diagonal, one easily sees that
the largest corrections to the matrices $(m^2_{(e,\nu)LL})_K$ and $(m^2_{eRR})_K$ come from electroweak gauge interactions and are proportional to the identity matrix in flavour space. Due to the negative sign of the dominant contribution these diagonal elements increase by evolving the mass matrices from the cutoff scale down to the electroweak scale. 

The largest corrections on the off-diagonal elements of the LL and RR low-energy soft mass are in the (13) and (23) sectors:
\beq
\begin{aligned}
(m^2_{(e,\nu)LL})_{K_{i3}}\Big|_{m_{susy}}&\approx \dfrac{1}{16 \pi^2}~ y_\tau^2\, (m^2_{(e,\nu)LL})_{K_{i3}}\log\left(\dfrac{\Lambda_L}{m_{SUSY}}\right)\,,\\
(m^2_{eRR})_{K_{i3}}\Big|_{m_{susy}}&\approx \dfrac{1}{16 \pi^2}~ y_\tau^2\, (m^2_{eRR})_{K_{i3}}\log\left(\dfrac{\Lambda_L}{m_{SUSY}}\right)\,.
\end{aligned}
\eeq
All such contributions can be safely neglected, because they are suppressed by $y_\tau$ and by $\log\left(\Lambda_L/m_{SUSY}\right)$.

The matrix $A_e$ gets a first correction by an overall multiplicative factor that can be absorbed, for instance, by a common
rescaling of the parameters, and a second correction of the type $A_e\to A_e+ \delta\, Y_e$, which is only into the diagonal elements and easily reabsorbable redefining the parameters. Finally, such corrections to the off-diagonal elements of $A_e$ are negligible.
We can conclude that the corrections on the off-diagonal entries of the soft mass matrices induced by the RG running are either negligible or could be absorbed into the parametrization.\\

When considering the See-Saw case, the RGEs that describe the running of the sfermion masses are different \cite{Martin:1993zk,Hisano:1995cp}:
\begin{align}
\begin{split}
\mu \dfrac{\mrm{d}}{\mrm{d}\mu}(m^2_{eLL})_{ij}=\mu \dfrac{\mrm{d}}{\mrm{d}\mu}(m^2_{eLL})_{ij}\Bigg|_{MSSM}+
\dfrac{1}{16\pi^2}\Bigg[\left(\hat m^2_{eLL}\hat Y_\nu^\dag \hat Y_\nu+\hat Y_\nu^\dag \hat Y_\nu m^2_{eLL}\right)_{ij}+\\
+2\left(\hat Y_\nu^\dag m^2_{\nu LL}\hat Y_\nu+m^2_{H_u}\hat Y_\nu^\dag \hat Y_\nu+ A_\nu^\dag A_\nu\right)_{ij}\Bigg]\,,
\end{split}
\label{RGEmLLRH}\\
&\mu \dfrac{\mrm{d}}{\mrm{d}\mu}(m^2_{eRR})_{ij}=\mu \dfrac{\mrm{d}}{\mrm{d}\mu}(m^2_{eRR})_{ij}\Bigg|_{MSSM}\\
&\mu \dfrac{\mrm{d}}{\mrm{d}\mu}(A_e)_{ij}=\mu \dfrac{\mrm{d}}{\mrm{d}\mu}(A_e)_{ij}\Bigg|_{MSSM}+\dfrac{1}{16\pi^2}\left[2\left(Y_e\hat Y_\nu^\dag A_\nu\right)_{ij}+\left(A_e\hat Y_\nu^\dag \hat Y_\nu\right)_{ij}\right]
\label{RGEYeRH}
\end{align} 
where $\hat Y_\nu$ is the neutrino Dirac Yukawa in the basis of diagonal charged leptons and RH neutrinos. Since two of the lepton mixing angles are large, then $\hat Y_\nu$ in general has large off-diagonal entries. This would contribute to flavour violating processes much more than in the case without RH neutrinos, especially from the $LL$ sector.

The low-energy quantities should be calculated considering the flavour boundary conditions at the scale $\Lambda_L\approx\Lambda_f$ and evolving them down by the use of eqs.~(\ref{RGEmLLRH})--(\ref{RGEYeRH}). This can easily be done numerically, but not analytically.


%

\providecommand{\WileyBibTextsc}{}
\let\textsc\WileyBibTextsc
\providecommand{\othercit}{}
\providecommand{\jr}[1]{#1}
\providecommand{\etal}{~et~al.}

\end{document}